\documentclass[acmtog,screen,nonacm,obeypunctuation]{acmart}

\usepackage[english]{babel}
\usepackage[shortcuts]{extdash}
\usepackage{hyperref}
\usepackage{tikz}
\usetikzlibrary{shapes,calc,positioning,arrows}
\pgfarrowsdeclarecombine{twotriang}{twotriang}
{stealth'}{stealth'}{stealth'}{stealth'}
\usepackage{mathtools}
\usepackage{microtype}
\usepackage{enumitem}
\usepackage{graphicx}
\usepackage{ifthen}
\usepackage[normalem]{ulem}
\usepackage[textwidth=14mm, textsize=tiny]{todonotes}

\usepackage{booktabs}
\usepackage{multirow}
\usepackage{tabularx}
\usepackage{colortbl}
\usepackage{threeparttable}
\usepackage{siunitx}
\usepackage{nicefrac}
\usepackage[noabbrev]{cleveref}
\usepackage{stackrel}

\definecolor{a}{HTML}{c0c0c0}
\definecolor{b}{HTML}{000000}
\definecolor{c}{HTML}{ffffff}
\definecolor{d}{HTML}{d36060}
\definecolor{e}{HTML}{80d380}
\definecolor{f}{HTML}{8080ff}
\definecolor{g}{HTML}{f0e462}
\definecolor{h}{HTML}{c49680}

\definecolor{i}{HTML}{c0c0c0}
\definecolor{j}{HTML}{000000}
\definecolor{k}{HTML}{ffffff}

\makeatletter
\newcommand\footnoteref[1]{\protected@xdef\@thefnmark{\ref{#1}}\@footnotemark}
\makeatother

\crefname{ineq}{inequality}{inequalities}
\crefname{subsection}{section}{sections}
\Crefname{subsection}{Section}{Sections}
\creflabelformat{ineq}{#2{\upshape(#1)}#3}

\authorsaddresses{}

\title{Dave: a decentralized, secure, and lively fraud-proof algorithm}
\author{Diego Nehab}
\affiliation{\institution{Cartesi Research}}
\author{Gabriel Coutinho de Paula}
\affiliation{\institution{Cartesi Research}}
\author{Augusto Teixeira}
\affiliation{\institution{Cartesi Research}} \affiliation{\institution{IMPA}}

\makeatletter
\newtheorem*{rep@theorem}{\rep@title}
\newcommand{\newreptheorem}[2]{%
\newenvironment{rep#1}[1]{%
 \def\rep@title{#2 \ref{##1}}%
 \begin{rep@theorem}}%
 {\end{rep@theorem}}}
\makeatother

\newreptheorem{theorem}{Theorem}

\newcommand{\nf}[2]{\ensuremath{\nicefrac{#1}{#2}}}

\makeatletter
\everymath{\if@display\else\medmuskip=2mu plus 1mu\thickmuskip=3mu plus 1.5mu\fi}
\makeatother

\makeatletter
\def\@secfont{\sffamily\Large\section@raggedright}
\makeatother

\renewcommand{\keywordsname}{Keywords}%

\renewcommand{\bibliofont}{\small}

\newcommand{\censorshipTimeBudget}{\ensuremath{T_c}}
\newcommand{\censorshipTimeAccumulated}{\ensuremath{T_d}}
\newcommand{\actionTimeSpent}{\ensuremath{T_a}}
\newcommand{\roundDuration}{\ensuremath{T_r}}
\newcommand{\groupSize}{\ensuremath{G}}
\newcommand{\nSybils}{\ensuremath{N}}
\newcommand{\maxDemotions}{\ensuremath{K}}
\newcommand{\stateLetter}{\ensuremath{S}}
\newcommand{\stateLetterPrime}{\ensuremath{\stateLetter^\prime}}
\newcommand{\stateAt}[2][\stateLetter]{\ensuremath{#1_{#2}}}
\newcommand{\statePrimeAt}[1]{\ensuremath{{\mathrlap{\stateLetterPrime}\phantom{\stateLetter}}_{#1}}}
\newcommand{\stateHashes}{\ensuremath{\mathbf{\stateLetter}}}
\newcommand{\stateHashesPrime}{\ensuremath{\mathbf{\stateLetterPrime}\!}}
\newcommand{\stateHash}[1]{\ensuremath{h(#1)}}
\newcommand{\compHash}[3]{H(#1, #2, #3)}
\newcommand{\actionDuration}[1]{\ensuremath{\tau_{#1}}}
\newcommand{\actionCount}{\ensuremath{I}}
\newcommand{\totalActionDuration}{\ensuremath{T_{\mathit{\mkern-2mu m}}}}
\newcommand{\gracePeriod}{\ensuremath{T_{\mathit{\mkern-2mu g}}}}
\newcommand{\demotions}[1]{\ensuremath{\boldsymbol{d}_{\mkern-2mu#1}}}
\newcommand{\borrowFirst}[2]{\ensuremath{\boldsymbol{a}}_{\mkern-2mu#1, #2}}
\newcommand{\borrowLast}[2]{\ensuremath{\boldsymbol{b}}_{\mkern-2mu#1, #2}}
\newcommand{\invariant}[2]{\ensuremath{\mathcal{S}_{#1,#2}}}
\newcommand{\invariantInc}[1]{\ensuremath{w_{#1}}}
\newcommand{\leastDemotions}[1]{\ensuremath{m_{#1}}}
\newcommand{\maxRoundDelay}[3]{\ensuremath{R_{#1,#2}(#3)}}
\newcommand{\nRounds}{\ensuremath{R}}
\newcommand{\maxDisputeTimeNoTc}{\ensuremath{\Delta T}}
\newcommand{\maxDisputeTimeTc}{\ensuremath{\Delta T'}}
\newcommand{\demotionCounter}{\ensuremath{k}}
\newcommand{\roundCounter}{\ensuremath{j}}
\newcommand{\transitionCounter}{\ensuremath{t}}
\DeclareMathOperator{\step}{step}

\newcommand{\aremark}[1]{\todo[color=cyan]{A\@: {#1}}}
\newcommand{\cremark}[1]{\todo[color=yellow]{C\@: {#1}}}

\newcommand{\lel}[1]{\ensuremath{\stackrel[#1]{}{\le}}}

\begin{abstract}
In this paper, we introduce a new fraud-proof algorithm that offers an unprecedented combination of decentralization, security, and liveness.
The resources that must be mobilized by an honest participant to defeat an adversary grow only logarithmically with what the adversary ultimately loses.
As a consequence, there is no need to introduce high bonds that prevent an adversary from creating too many Sybils.
This makes the system very inclusive and frees participants from having to pool resources among themselves to engage the protocol.
Finally, the maximum delay to finalization also grows only logarithmically with total adversarial expenditure, with the smallest multiplicative factor to date.
In summary: the entire dispute completes in 2--5 challenge periods, the only way to break consensus is to censor the honest party for more than one challenge period, and the costs of engaging in the dispute are minimal.
\end{abstract}

\date{november}

\begin{document}

\maketitle

\section{Introduction}

\emph{Fraud-proof} algorithms allow a computationally limited referee to identify, between a set of competing players, the honest claim concerning the result of a long computation~\citep{Feige1997,CanettiRivaEtAl2013}.

The key idea is as follows.
Referee and players must agree on a deterministic \emph{state transition function} and on an \emph{initial state}.
The players first compute the \emph{final state} by successively applying the state transition function over the initial state.
They then make a \emph{claim} by submitting their results to the referee.
If all players agree, the referee accepts the result as correct.
Otherwise, the referee guides the players into a binary search that progressively \emph{bisects} the computation to identify the earliest disputed state transition.
Finally, by verifying this single state transition, the referee can identify and eliminate a dishonest player.

In the context of programmable blockchains like Ethereum~\citep{wood2024ethereum}, fraud proofs play a central role in \emph{optimistic rollups}~\citep{Rollups, rollup-roadmap}.
Here, the computationally limited blockchain acts as the referee, adjudicating disputes between external validators~\citep{TeutschEtAl2017, Arbitrum2018, TeixeiraNehab2018}.
The security of such a system hinges on two properties.
First, a single honest validator must prevail over any number of dishonest ones and enforce the final state.
Second, anyone must be able to become a validator.
In other words, securing any optimistic rollup should be just a matter of setting up a node to run on your own infrastructure.
This allows an optimistic rollup to inherit the security of Ethereum.

Unfortunately, allowing just anyone to join the validator set opens the floodgates for an attacker to create as many Sybils as it has resources for~\citep{douceur2002sybil}.
Sybil attacks are hard to defeat~\citep{fraud-proofs-broken, DelayAttacks}.
It is quite a challenge (see \cite{fraud-proof-war}) to design protocols where defeating any single honest validator requires an attacker to spend an impractical quantity of resources (\emph{security}), that do not restrict participation by imposing a permission list or demanding the deposit of expensive \emph{bonds} (\emph{decentralization}), and that can quickly settle disputes (\emph{liveness}).

In this paper, we introduce \emph{Dave}, a new fraud-proof algorithm that is resilient to Sybil attacks.
Dave allows (but does not require) honest validators to cooperate trustlessly, playing collectively as one.
We call this set of honest players that can fight together the \emph{hero}.
We call the attacker and its mob of perfectly coordinated Sybils the \emph{adversary}.

Launching a Sybil attack in Dave is exponentially expensive for the adversary, relative to the resources the hero must allocate to prevail.
In practice, this enables a singleton hero to defeat a nation-state adversary mounting any realistic Sybil attack, even with cheap bonds and no permission lists.
The hero needs a constant amount of hardware to participate in Dave, regardless of the number of Sybils, and the costs it incurs are reimbursed after the dispute is over.

Crucially, settlement delay grows logarithmically on the number of Sybils, but with a key advantage over the work of~\citet{nehab2022permissionless}: the constant multiplying the logarithm is one order of magnitude smaller.
Like other optimistic rollups, Dave requires at least one \emph{challenge period}\footnote{In Ethereum, due to censorship considerations, the challenge period is taken to be one week (see \cref{threat}).} to settle.
However, Dave settles disputes within 2--5 challenge periods for any realistic number of Sybils.

\section{Related work}\label{related}

In theory, rollup solutions should prefer \emph{validity proofs}~\citep{groth2016size,ben2018scalable} over fraud proofs.
A validity proof allows the computationally limited referee, when presented with ancillary information (a \emph{witness}), to directly verify the result of a computation with little effort.
There is no challenge period or dispute resolution: results are final, ``immediately''.
Furthermore, the effort to create a proof needs to be spent only once.
Rollups based on validity proofs are called \emph{ZK rollups}.

Unfortunately, current algorithms for producing the required witnesses are many orders of magnitude more computationally expensive than simply running the computation itself.
More precisely, to generate the witnesses for a computation that can be comfortably processed in an ordinary laptop, a validity-proof system would require a supercomputer.
This has two negative effects: it drastically decreases throughput, and it creates a centralizing effect on the liveness of the network (see \cite{zk-cost}).
Nevertheless, in practice, validity proofs can be a valuable tool for several applications, including when designing fraud-proof algorithms (see~\cref{computation-hash}).

Fraud-proof algorithms allow for significantly greater throughput on affordable hardware.
In the subsequent discussion, we will analyze existing fraud-proof algorithms from the perspective of security, decentralization, and liveness, as these criteria are a revealing lens through which the design choices of competing fraud-proof algorithms can be contrasted.

Decentralization has a complex interplay with security and liveness.
Indeed, protocols often favor one to the detriment of another.
For example, liveness and security may depend on preventing the adversary from creating too many Sybils.
A possible mechanism is to increase bond values.
This effectively harms decentralization by indiscriminately restricting the number of players that can afford to participate.
Another approach is running disputes in parallel, which may considerably improve liveness.
However, it allows an attacker to drown the honest players in a sea of Sybils, severely harming security.

The fundamental fraud-proof idea used in optimistic rollups (state hash as claim, followed by bisection, then execution of a single step) was first proposed by~\citet{CanettiRivaEtAl2013}.
In its original formulation, claims must be identified with individual players who must personally defend them.
Otherwise, a dishonest player could present an honest claim and then misrepresent its defense to eliminate it.

\citeauthor{CanettiRivaEtAl2013}\ propose two alternatives for disputes involving multiple players/claims.
The first suggestion is a sequence of rounds, in which all players participate and one dishonest player is eliminated per round.
The second proposal runs in parallel and finishes in a single round, wherein every player joins a separate simultaneous match against every other player that made an incompatible claim.

When many players are present (i.e.\ a Sybil attack), the sequential algorithm can produce indefinite settlement delays (no liveness), whereas the parallel algorithm demands unbounded resources from the honest players (no security).
The alternative is to limit the maximum number of players (no decentralization).

PRT~\citep{nehab2022permissionless} introduces a stronger type of claim called the \emph{computation hash}, which is essentially the hash of the entire sequence of state hashes throughout the computation (see~\cref{computation-hash}).
Computation hashes obviate the association between players and claims because referees can identify and outright reject any attempted perjury committed in the defense of a claim.
Honest players can collaborate without trusting each other, forming the hero.
Making a claim requires posting a bond, but bisections and the final step do not.

Since dense computation hashes (as originally proposed) are expensive to calculate, PRT uses sparse computation hashes as claims.
All unique claims are arranged into a tournament bracket.
Matches between sparse computation hashes must be refined into sub-tournaments based on progressively denser computation hashes that cover progressively smaller parts of the computation.
In such a multi-stage dispute, the adversary's bond expenditure affects the hero in a poly-logarithmic way.
The facts that bonds can be relatively cheap, that the hero only needs as many bonds as there are levels in the tournament, and that the hero is involved in a single match per round, allow the hero to spend few resources (decentralization).
The poly-logarithmic impact prevents the adversary from exhausting the hero's resources, even if there is a single independent honest player (security).
The problem is that, since the adversary's censorship budget is recharged after each tournament round, the censorship budget multiplies the poly-logarithmic factor, leading to large settlement delays (poor liveness).

Like PRT, BoLD~\cite{alvarez2024bold} uses sparse computation hashes as claims.
Unlike PRT, the claims are not arranged into a tournament bracket.
Instead, the hero must concurrently defeat all dishonest claims and protect the honest claim against all challenges.
It can be understood as the combination between the parallel approach of \citet{CanettiRivaEtAl2013} with the stronger claim of PRT\@.
This combination leads to optimal liveness: disputes complete within a single additional challenge period.
To be secure under current parameters, the multi-stage dispute would require the hero to invest at least 13.4\% of the value the adversary is willing to burn, \citep{Arbitrum-Economics, fraud-proof-war}.
In other words, if the adversary is 6.5 times better funded than the hero, the adversary will be able to exhaust the hero's resources and win (stressing security).
BoLD requires expensive bonds and parallel engagement, which excludes independent players (poor decentralization).
It is important to note that there is an incentive for well-funded attacks against BoLD because they have the potential to capture all the value locked inside the chain.
In contrast, this vulnerability (and therefore the incentive) is \emph{not} present in Dave, where a well-funded attack can only slightly increase the delay to finalization.

OP~\citep{Optimism_2024} uses regular state hashes as claims, instead of computation hashes.
Nevertheless, claims are not identified with players.
Anyone can bisect any claim, but all interactions require the posting of a bond: claims, bisections, and the step.
Like in BoLD, the hero must defeat all dishonest claims and protect the honest claim against all challenges, in parallel.
To be secure, the hero would need at its disposal at least 51\% of the value the adversary is willing to burn~\citep{fraud-proof-war}.
These requirements undermine the security and decentralization of the method.

The algorithm we present in this paper is an evolution over PRT\@.
As in PRT, claims are computation hashes, and bonds are only required when posting a claim.
A parameter to the algorithm controls the maximum number of matches the hero is involved in simultaneously, per round (typically, 1, 3, or 7).
This can buy improved liveness by increasing computational demands on the hero.
Dave replaces PRT's fixed tournament bracket with a dynamic matching strategy.
Crucially, the new matching strategy does not require the adversary's censorship budget to be recharged after each round, allowing Dave to amortize it over the entire dispute.
Moreover, Dave uses validity proofs to verify a relatively ``fat'' step at the end of the bisection, but not while generating the computation hash.
This enables dense computation hashes (i.e., a single-stage dispute) without the normally associated overhead of generating it.
Because of these changes, the poly-logarithmic factor becomes a simple logarithm, and the constant multiplying it is one order of magnitude smaller when compared to that of PRT\@.
Recall PRT was already decentralized and secure.
Dave maintains these properties, while bringing the desired liveness.

The same technique that allows Dave to use dense computation hashes could also be applied to PRT and BoLD\@.
In the case of PRT, the single-stage tournament would transform the poly-logarithm into a simple logarithm.
However, the censorship budget remains as the constant that multiplies it, eroding liveness.
In the case of BoLD, the resulting single-stage dispute would lower the hero's investment to 0.1\% of the value committed by the adversary.
Nevertheless, high bond prices remain as a barrier to decentralization and the parallel dispute would mandate access to extraordinary computational power for producing the required witnesses.

\section{Threat and resource model}\label{threat}

In optimistic rollups, the referee is implemented as a set of blockchain smart contracts.
As a consequence, its security depends on certain assumptions concerning the way blockchains operate.
We assume that all transactions are processed correctly, eventually.
However, we assume the adversary can subvert this process, to a limited extent, by:

\begin{description}[leftmargin=1.5em]
  \item[\normalfont\emph{Censoring transactions}]
  The adversary has the power to delay any set of transactions.
  The only limitation to this power is that the total amount of time during which it is exerted cannot not exceed the \emph{censorship time budget}~$\censorshipTimeBudget$, usually taken to be a week in Ethereum~\citep{kelvinfichter_challenge}.
  This means the adversary may choose to use up all its budget in one go and delay any set of transactions for a week, or partition its budget of $\censorshipTimeBudget$ in multiple spans of censorship.
  This is why we \big(and others~\citep{Arbitrum2018, alvarez2024bold}\big) set the \emph{challenge period} to the censorship time budget~$\censorshipTimeBudget$;
  \item[\normalfont\emph{Reordering transactions}]
  The adversary has the power to reorder incoming transactions on the blockchain, for example by front\-/running honest validators.
\end{description}

Interactive fraud-proof protocols require players to take turns when interacting with the referee.
If no deadlines are set, the adversary could stall the dispute indefinitely.
Conversely, the adversary could use its censorship budget to force the hero to miss a deadline.
Fraud-proof algorithms must therefore take the censorship budget into account when setting the penalties incurred for missing any deadline.

It is quite a challenge to design lively protocols around a high $\censorshipTimeBudget$, and $1$ week is very high indeed.
Pessimistically setting the deadline of every interaction to at least $\censorshipTimeBudget$ would ruin the liveness of any protocol.
The protocols discussed in~\cref{related} use strategies like chess clocks to amortize $\censorshipTimeBudget$ over many interactions.
See~\cref{dave} for a novel approach that amortizes~$\censorshipTimeBudget$ over the entire dispute.

Players interact by making \emph{moves} (see~\cref{match}), which require spending resources.
We model three kinds of resources, that we assume are priced so there is an exchange rate between them:

\begin{description}[leftmargin=1.5em]
  \item[\normalfont\emph{Gas}]
    All moves add a transaction to the base layer, which requires paying a transaction fee in base layer tokens.
    If the hero runs out of gas, it will no longer be able to participate in the dispute;
  \item[\normalfont\emph{Compute}]
    In order to react correctly, players may need to run costly computations.
    These computations may exhaust the hero's compute capacity, and force it to forfeit the dispute;
  \item[\normalfont\emph{Bonds}]
    As a protection against Sybil attacks, some moves require depositing tokens into the protocol itself, as a collateral.
    In this sense, bonds are endogenous, and their cost is arbitrarily set by the protocol.
\end{description}

The only hope an adversary has to defeat the hero is to exhaust its resources, preventing it from taking part in the dispute and compromising the security of the algorithm.
Because of this, it is not sufficient to consider the hero as the total number of honest agents.
The analysis must include the resources held by the hero as well.
We need a more nuanced understanding of what ``one honest validator'' means.

For that, we take inspiration from~\citet{Buterin_2020}.
He describes trust as the use of any assumptions about the behavior of other people, classifying trust models as ``the ability of the application to continue operating in an expected way without needing to rely on a specific actor to behave in a specific way''.
When an algorithm allows $N$~players to participate but requires $M$~of~them to be honest, he argues, $M$ should be as small as possible (ideally 1) and $N$ as large as possible.

This maps naturally to what we have described as security and decentralization.
In the context of fraud proofs, decentralization demands an arbitrarily large value for~$N$, and security demands---\emph{in addition}---that $M$ be $1$.
Algorithms that have these two properties (i.e., a single honest validator can enforce the correct result, and that validator can be anybody) inherit the security of the base layer.
Through this lens, we define ``one honest player'' not as an individual agent, but rather as a unit of a total of $N$~resources, $M$ of which are spent to defend correct behavior\footnote{Note that when the security of an algorithm requires a constant ratio $M/N$, there are two alternatives.
Either $M$ and $N$ are both kept small (harming decentralization), or honest players will be forced to marshal a lot of resources (harming security).}.

We assume the existence of at least one honest player.

\section{Dave}\label{dave}

Here is an informal overview of a dispute resolution under Dave:
\begin{enumerate}
  \item The hero submits the honest claim, and each Sybil submits a dishonest one.
    Claims are unique.
    All competing claims must be posted within~\censorshipTimeBudget.
    Rounds start as soon as there is more than one competing claim;
  \item\label{itm:match-item}
    At the start of a round, Dave partitions all surviving claims into small groups of size~\groupSize.
    Each group spawns one match for every pair of claims within.
    All matches of all groups in a round are concurrent, starting together and ending together.
    The round duration~\roundDuration\ is fixed and is significantly shorter than~\censorshipTimeBudget;
  \item
    A match concludes with one claim defeating the other, either by timeout or through a step action.
    The claim that wins all matches wins its group, and all the other claims lose;
  \item
    Dave doesn't eliminate claims outright.
    Instead, it keeps a tally of the number of times each claim has failed to win its group.
    When a claim has failed \emph{enough times} to ensure censorship cannot be blamed, it is eliminated;
  \item
    After at least \censorshipTimeBudget\ has elapsed, if there is a single surviving claim, it wins the dispute.
    Otherwise, repeat from~\cref{itm:match-item}.
\end{enumerate}

For completeness, we start by describing the computation hash as claim (see~\cref{computation-hash}), and how it enables trustless collaboration.
We then describe what happens during a match (see~\cref{match}).
We proceed to describe how our chess clock treats censorship differently from the chess clock of~\citet{Arbitrum2018} (see~\cref{clock}).
We complete the algorithm by describing the novel matchmaking strategy at the heart of Dave (see~\cref{matchmaking}).
%

\subsection{Computation model}\label{computation-model}

The referee and players must agree on an \emph{initial state}~\stateAt{0}\ and on a \emph{state transition function}~$\step$.
The state at \emph{time} \transitionCounter\ is $\stateAt{\transitionCounter} = \step(\stateAt{\transitionCounter-1})$.
It is obtained from the~\transitionCounter{-th} iteration of the state transition function over the initial state.
We assume the computation terminates after a power~of~two number of steps, so the \emph{final state} is~\stateAt{2^B}.
There is no loss of generality since, if the computation completes before~$\transitionCounter'$, then the state transition has the property that~$\stateAt{\transitionCounter'} = \step(\stateAt{\transitionCounter'})$.
In other words: the final state is a fixed point of~$\step$.
As far as we are concerned, the state is opaque.
All we need is that~\stateAt{\transitionCounter}\ can be identified by a collision\-/resistant \emph{state hash} $\stateHash{\stateAt{\transitionCounter}}$ and that the referee can verify a state transition given a pair of candidate consecutive state hashes.

\subsection{Structure of a claim}\label{computation-hash}

In the work of~\citet{CanettiRivaEtAl2013}, a claim is simply~$\stateHash{\stateAt{2^B}}$, the final state hash.
Two disagreeing players engage in an interactive binary search to pinpoint the time at which they first disagreed (recall they agree on the initial state hash).
Each interaction is a bisection that cuts the search space by half.

This setup is susceptible to Sybil attacks in which a dishonest player creates many false claims.
This forces the hero to play an equally large number of matches.
Either these matches are played in parallel, which could exhaust the hero's resources, or the matches are in series, which can cause large settlement delays.

A first impulse in solving this problem is to force Sybils to fight and eliminate one another, effectively helping the hero to prune dishonest claims out.
Unfortunately, this idea is vulnerable to false flag attacks: the adversary can claim the correct final state but lose the game on purpose by misrepresenting its defense (i.e., by performing the bisection incorrectly).

\citet{nehab2022permissionless} introduced the \emph{computation hash}, a stronger claim that resists misrepresentation.
Rather than commit to just the state, computation hashes commit to the entire computation history.
A similar technique was then adopted by BoLD.

Informally, a computation hash is the root hash of a binary Merkle tree whose leaves are $\stateHashes = \big[\stateHash{\stateAt{1}}, \dots, \stateHash{\stateAt{2^B}}\big]$, the state hashes \emph{after} every state transition in the history of the computation (therefore excluding the agreed-upon $\stateHash{\stateAt{0}}$).

\begin{figure}
  \begin{tikzpicture}[<-, >=stealth', auto,
    level/.style={sibling distance=60mm/#1, font=\footnotesize}, scale=.55,
    level 4/.style={sibling distance=22mm}
    ]
    \node [rectangle,draw] (z){$\compHash{\stateHashes}{0}{B}$}
    child {node [rectangle,draw] (a) {$\compHash{\stateHashes}{0}{B-1}$}
      child {node [rectangle,draw] (b) {$\compHash{\stateHashes}{0}{B-2}$}
        child {node {$\vdots$}
          child {node [rectangle,draw] (d) {$\stateHash{\stateAt{1}}$}}
          child {node [rectangle,draw] (e) {$\stateHash{\stateAt{2}}$}}
        }
        child {node {$\vdots$}}
      }
      child {node [rectangle,draw] (g) {$\compHash{\stateHashes}{1}{B-2}$}
        child {node {$\vdots$}}
        child {node {$\vdots$}}
      }
    }
    child {node [rectangle,draw] (j) {$\compHash{\stateHashes}{1}{B-1}$}
      child {node [rectangle,draw] (k) {$\compHash{\stateHashes}{2}{B-2}$}
        child {node {$\vdots$}}
        child {node {$\vdots$}}
      }
      child {node [rectangle,draw] (l) {$\compHash{\stateHashes}{3}{B-2}$}
        child {node {$\vdots$}}
        child {node (c){$\vdots$}
          child {node [rectangle,draw] (o) {$\stateHash{\stateAt{2^B-1}}$}}
          child {node [rectangle,draw] (p) {$\stateHash{\stateAt{2^B}}$}
          }
        }
      }
    };
  \end{tikzpicture}
\vspace*{-2ex}
  \caption{The construction of the computation hash as a Merkle tree.}\label{fig:comphashtree}
\vspace*{-3ex}
\end{figure}
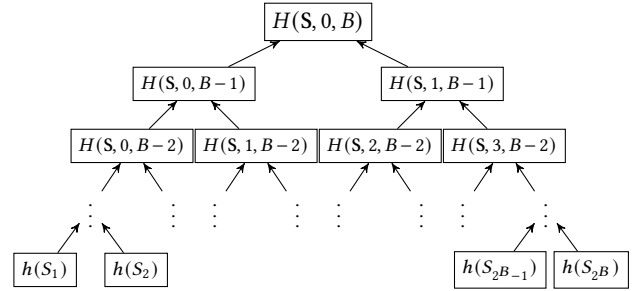

More precisely, for the $i$th subtree of height $b$ in~\cref{fig:comphashtree}, we define the computation hash $\compHash{\stateHashes}{i}{b}$.
In the end of this construction, $\compHash{\stateHashes}{0}{B}$ will be the desired computation hash.
Starting with the leaves ($b = 0$), we define $\compHash{\stateHashes}{i}{0} = \stateHash{\stateAt{i+1}}$, which is the $i$th element in \stateHashes, and \mbox{$(i+1)$-th} state in the overall computation (recall that \stateHashes\ doesn't contain \stateAt{0}).
The computation hash~$\compHash{\stateHashes}{i}{b}$ of height $b$ is constructed by hashing the concatenation of its left and right children $\compHash{\stateHashes}{2 i}{b-1}$ and~$\compHash{\stateHashes}{2 i + 1}{b-1}$, respectively.

We now require that every bisection be accompanied by a valid proof that it is consistent with the computation hash in the claim (see~\cref{match} for further details).
The advantage is that, once committed, players can no longer lie during bisection.
In particular, an adversary that posts the honest computation hash cannot single-handedly lose the game on purpose.
There is a fundamental shift: computation hashes reveal lies instead of just liars, turning disputes into contests between claims rather than between individual players.
Consequently, Dave can safely allow any player to represent any claim at any time.
Players' signatures need no longer be part of the message, and the set of honest validators can trustlessly join forces.

When the bisection has narrowed down disagreement to a single application of the state transition function, the referee must be able to identify the state hash that follows the last agreed-upon state hash in the computation hash.
Recall the referee runs on the blockchain.
If it is to execute the state transition function directly, then this must be a rather inexpensive one (e.g., a single instruction of a simple virtual machine).
Unfortunately, updating the state hash is several orders of magnitude more expensive than executing such a simple state transition.
Therefore, performing this update after every state transition makes the overhead of computing dense computation hashes impractical in this context.
This forced PRT and BoLD to amortize the cost of computing a state hash over the cost of many state transitions, by using \emph{sparse computation hashes} as claims.
These record state hashes periodically, once every many state transitions.
As we discuss in~\cref{related}, this sparsity requires multi-level disputes, which are problematic.

When the computational cost of the state transition function is comparable to the cost of updating the state hash, \emph{dense computation hashes} that record a state hash after every state transition become practical.
To that end, Dave uses a relatively ``fat'' state transition (i.e., corresponding to many instructions of a complex virtual machine).
The trick is that the referee never executes the state transition itself.
Instead, after the bisection stage, the referee demands a witness for (and verifies) a validity proof that a given state hash follows the last agreed-upon state hash.
Crucially, in the process of creating a computation hash, no validity-proof witnesses must be generated or verified.
While producing a claim, the virtual machine can execute its ``fat'' state transitions at native speeds between state-hash updates.
This is many orders of magnitude faster than stopping after every virtual-machine instruction to update the state hash, and many additional orders of magnitude faster than creating a validity proof for the entirety of the computation.
Finally, the hero only needs to produce a single validity proof per match, each covering a very small part of the entire computation (i.e., a single ``fat'' state transition), and the number of simultaneous matches on which the hero is involved is limited by the small group size.

\subsection{Match between two claims}\label{match}

Recall that, in the beginning of every round, Dave partitions all surviving claims into small groups with at most \groupSize\ claims.
Each claim then plays an independent, simultaneous match against every other claim in its group.

Consider two computation hashes paired in a match, $C = \compHash{\stateHashes}{0}{B}$ and~\mbox{$C^\prime\!= \compHash{\stateHashesPrime\!}{0}{B}$}.
These computation hashes implicitly agree on~\stateAt{0}.
However, their computation histories~\stateHashes\ and~\stateHashesPrime\ disagree in at least one of their states.
Validators of $C$ and~$C^\prime$ must perform a sequence of \emph{actions} to try to refute the opponent's claim.

A match progresses by first isolating the divergence between $C$ and~$C^\prime$ to the first state in which they disagree.
In other words, by finding the smallest \transitionCounter\ such that \mbox{$\stateAt{\transitionCounter} = \statePrimeAt{\transitionCounter}$} and~\mbox{$\stateAt{\transitionCounter+1} \neq \statePrimeAt{\transitionCounter+1}$}.
Then, by verifying a single state transition, Dave can eliminate one of the two claims.
It can eliminate~$C$ if $\statePrimeAt{\transitionCounter+1} = \step(\statePrimeAt{\transitionCounter})$ or~$C^\prime$ if $\stateAt{\transitionCounter+1} = \step(\stateAt{\transitionCounter})$.
If a validator does not act in a timely manner, the match ends in a timeout (see~\cref{clock}).

There are two kinds of action in a match:
\begin{description}[leftmargin=1.5em]
\item[\normalfont\emph{Bisect}]
  The validators of $C$ and $C^\prime$ take \emph{bisect} actions in turns.
  A match starts at turn $b = B$, and progresses down to $b = 0$.
  At the beginning, $C$ and $C^\prime$ are known to disagree in the computation hash $\compHash{\stateHashes}{0}{B} \neq \compHash{\stateHashesPrime\!}{0}{B}$.
  At every subsequent turn $b - 1$, $C$ and $C^\prime$ are known to disagree in some subtree~$i$ at height~\mbox{$b - 1$}, meaning $\compHash{\stateHashes}{i}{b-1} \neq \compHash{\stateHashesPrime\!}{i}{b-1}$.
  When bisecting, if this disagreement is shown to be on the left subtree, Dave narrows the disagreement down to $\compHash{\stateHashes}{2i}{b-1} \neq \compHash{\stateHashesPrime\!}{2i}{b-1}$.
  Otherwise, Dave narrows the disagreement down to $\compHash{\stateHashes}{2i+1}{b-1} \neq \compHash{\stateHashesPrime\!}{2i+1}{b-1}$.
    When $b$ reaches $0$, Dave has found the first diverging leaf $\compHash{\stateHashes}{i}{0} = \stateHash{\stateAt{i+1}}$, and implicitly the last agreeing state \stateAt{i}\ (which can potentially be \stateAt{0});

\item[\normalfont\emph{Step}]
  Once the divergence is found, either party can take a step action, submitting the witness needed for the referee to verify a state transition.
  Upon receiving the witness, Dave advances the last agreed-upon state and uncovers a wrong claim.
  This wrong claim forfeits the match, ending it.
\end{description}

There are two ways a match can end: by step action or by timeout.
It is safe to outright eliminate from the dispute any claims that lose through a step action.
This is because the hero can never lose a match by taking a step action, and neither can the adversary inappropriately win one against the hero in this way.
However, although this elimination could in practice shorten some disputes, it does not improve the worst-case analysis in~\cref{liveness-proof}.
As such, for the sake of simplicity, we make both the step action and timeouts have the same result: forfeiting just the that particular match.

\subsection{Chess clock}\label{clock}

At each turn in a match, a validator defending one of the claims is expected to act.
Ultimately, these actions require including a transaction in the blockchain.
There are different reasons for which a player might fail to submit bisect or step actions on time---Sybils may refuse to take actions, or honest participants may be censored.
In these situations, we introduce a timeout mechanism that causes the unresponsive claim to forfeit the match.

The timeout mechanism considers the time it takes to complete actions.
Let \actionDuration{i}\ be a pessimistic estimate the duration of action~$i$, and let $\totalActionDuration = \sum_{i=1}^\actionCount \actionDuration{i}$ be the total time for all~\actionCount~interactions of both players.
(In Ethereum, \totalActionDuration\ is taken to be a few hours, which is significantly smaller than the censorship time budget~\censorshipTimeBudget.)
We will consider progressively more realistic scenarios until we reach Dave's clock.

First, consider a no-censorship regime (i.e., $\censorshipTimeBudget = 0$), groups with only 2 claims (i.e., $\groupSize=2$), and a total of~\nSybils~Sybils.
The optimal strategy would set the timeout for the~\mbox{$i$th} action to its duration~\actionDuration{i}.
Validators would be expected to complete their action within that time window.
Any unresponsive claim would be eliminated from the entire dispute.
Under this scenario, each match would take at most \totalActionDuration, and result in a Sybil elimination.
Since all claims are matched pairwise and each match eliminates one Sybil, half the Sybils would be eliminated after every round.
This means there would be $\log_2\nSybils$ rounds before the hero was declared victorious, with a total duration of $\,\totalActionDuration\lceil\,\log_2\nSybils\,\rceil$.

If the adversary has any censorship power, this idea breaks down, as the hero would be unfairly eliminated if censored by any amount.
Any timeout strategy must take~\mbox{$\censorshipTimeBudget>0$} into consideration.
Until~\censorshipTimeBudget\ has elapsed, the referee has no way of knowing whether the inactivity is ``honest'' (the hero is being censored) or ``dishonest'' (a Sybil is refusing to engage).

The naive solution is to set the time window to~\mbox{$\censorshipTimeBudget + \actionDuration{i}$}.
After all, the adversary could decide to use up all its censorship budget in one go.
The total dispute duration would be~\mbox{$\big(\tfrac{1}{2} \actionCount\, \censorshipTimeBudget + \totalActionDuration\big) \lceil\,\log_2\nSybils\,\rceil$} in this case.
Unfortunately, since~\censorshipTimeBudget\ in Ethereum is taken to be one week, this strategy introduces unreasonable delays.

Using a clock like the ones used in chess matches allows for amortizing the cost of \censorshipTimeBudget\ over all actions taken throughout a match.
The idea is simple: give both claims an initial time budget of~\mbox{$\censorshipTimeBudget + \tfrac{1}{2} \totalActionDuration$}, and whenever one side is expected to act, keep its clock ticking.
Once one of the claims exhausts their time budget, it is eliminated by timeout.
In this setup, a match takes at most~\mbox{$2\censorshipTimeBudget + \totalActionDuration$}.
This is a remarkable improvement in the total match time in Ethereum, from several dozen weeks to one week plus a few hours.
However, the dispute still takes~\mbox{$(2\censorshipTimeBudget + \totalActionDuration)\lceil\,\log_2\nSybils\,\rceil$}, which can still be too long.

Dave introduces a novel technique to amortize~\censorshipTimeBudget\ over the whole dispute, instead of over a single match.
First, we introduce the \emph{grace period}~\gracePeriod.
Matches still use a chess clock, but they start with $\tfrac{1}{2}\totalActionDuration + \gracePeriod$.
Crucially, the hero can only lose a match if the adversary burns at least~\gracePeriod\ from its censorship budget~\censorshipTimeBudget.

Recall that, during a round, each a claim engages in independent, simultaneous matches against all other claims in its group.
All matches in a round start and end at the same time.
Accordingly, we set the \emph{round duration} to $\roundDuration = \totalActionDuration + 2\gracePeriod$.

When a claim loses its group due to a timeout, it cannot be eliminated outright.
This is because the referee cannot distinguish between a case of the hero being censored by the adversary while trying to defend the honest claim, and a case of an adversary's Sybils deliberately refusing to act while defending a bogus claim.
Fortunately, we know the hero can't be forced to timeout in more than~\mbox{${\censorshipTimeBudget}/\gracePeriod$} separate rounds.
This is because the adversary must burn~\gracePeriod\ from its censorship budget in the process, and it only has~\censorshipTimeBudget available in total.
Dave therefore keeps a \emph{demotion count} for each claim.
Every claim starts the dispute with zero demotions.
To preserve its demotion count after a round, a claim must win all matches in its group.
Otherwise, its demotion count is incremented by one.
A claim is finally eliminated when its demotion count reaches~$\maxDemotions$.
It is safe to do so because, under our threat model, the only possible explanation for these many demotions is a Sybil repeatedly running down its clock on purpose.

In conclusion, the hero may be demoted a few times due to censorship, but it will not lose the dispute.
Moreover, the dispute always progresses because, at every round, about half $\big(\text{or rather, }(\groupSize-1)/\groupSize\big)$ of the dishonest claims get closer to being eliminated.
Given enough rounds, the dispute will finish.
In subsequent sections, we discuss how long this process takes.

\subsection{Partitioning claims into groups}\label{matchmaking}

\begin{figure}
\includegraphics[width=\linewidth]{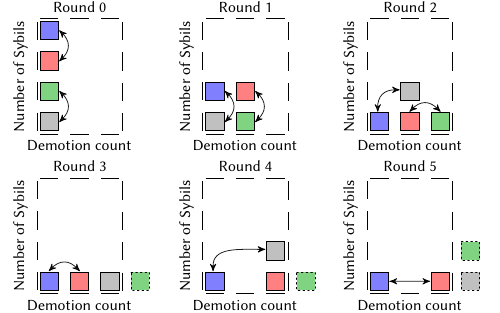}
\caption{First 5 rounds in a possible dispute with~$\nSybils=4$, $\maxDemotions=3$, and $\groupSize=2$.
Arrows show the groups before each round.}
\label{fig:rounds}
\end{figure}

After each round, all surviving claims are repartitioned into new groups in a process we call \emph{matchmaking}.
Dave's matchmaking is straightforward.
It sorts all claims by increasing demotion count and consumes them in order, forming groups of~\groupSize~claims each.
In this way, claims in each group tend to have similar demotion counts, to the best extent possible.
Naturally, the last group may have fewer than~\groupSize~claims.
Should a claim find itself alone in this last group, it is assumed to have won the group.
\Cref{fig:rounds} illustrates the process as the first 5~rounds in a dispute between 4~claims unfolds, using groups with 2~claims each.

To understand why it is important to deny the adversary control over the matchmaking process, let us run a thought experiment that assumes otherwise.
For simplicity, take~\mbox{$\groupSize=2$} during the experiment.
By controlling the matchmaking process, the adversary would first divide its Sybils in two halves.
It would sacrifice one half to spare the other half, by grouping a sacrificial Sybil with a to-be-spared Sybil, and making the sacrificial Sybil lose its group.
After~\maxDemotions~rounds, half the Sybils would have been eliminated from the dispute, while the other half would have kept their demotion count unscathed.
Since the adversary would lose half its remaining Sybils after every round, it would take it~\mbox{$\log_2\nSybils$} rounds to lose all Sybils.
Recall each round takes~\mbox{${\censorshipTimeBudget} + {\roundDuration}$}.
This strategy shows that control over matchmaking allows the adversary to delay settlement by~\mbox{$(\censorshipTimeBudget + \roundDuration)\log_2\nSybils$}~rounds.
This is no better than PRT~\citep{nehab2022permissionless}.
Taking the concrete value of these constants on Ethereum, it's feasible for a well-funded adversary to delay finality for several months (see~\cref{tight} for a discussion about these constants).
Dave's matchmaking was designed to avoid exactly this type of attack.

Note that the adversary still has some freedom under Dave's matchmaking.
Although Sybils with the same demotion count are fungible, some groups may contain claims with distinct demotion counts.
This is because the number of claims with each demotion count is not always divisible by~$\groupSize$.
In such cases, the adversary can select which Sybil will win each group, based on their demotion count.
As we shall see, this freedom is not enough for the adversary to mount an effective attack.

\section{Worst-case delay in rounds}

In~\cref{asymptotic}, we give an intuition for why the number of rounds Dave takes to settle a dispute grows only logarithmically with the number of Sybils.
The formal proof, presented in~\cref{liveness-proof}, is rather involved.
Then, in~\cref{strategy}, we explore computationally the space of adversary strategies to identify a maximum-delay strategy.
Finally, in~\cref{tight}, we analyze, again computationally, the maximum delay that can be achieved by the strategy just identified.
We offer compelling evidence that the liveness is significantly better than what the formal asymptotic bound guarantees.

\subsection{Asymptotic bound on worst-case delay}\label{asymptotic}

The following is an outline for the proof of~\cref{liveness-theorem}.
The formal proof, which unfortunately is quite involved, is found in~\cref{liveness-proof}.
\begin{theorem}\label{liveness-theorem}
When groups have at most~\groupSize~claims, and a claim can be demoted \maxDemotions~times before being eliminated, Dave settles a dispute involving \nSybils~claims in $O\big(\maxDemotions + \log_\groupSize \nSybils + \sqrt{\maxDemotions \log_\groupSize\nSybils}\,\big)$ rounds.
\end{theorem}

Let~$\demotions{\roundCounter}[\demotionCounter]$ denote the number of claims demoted exactly~$\demotionCounter$ times after~\mbox{$\roundCounter = 0, 1, \dots$}~rounds.
The proof in~\cref{liveness-proof} can be roughly divided into two parts.

In the first part, we show that, despite the adversary's best efforts, after only $J=O\big(\maxDemotions + \log_\groupSize \nSybils + \sqrt{\maxDemotions \log_\groupSize\nSybils}\,\big)$ rounds, it holds that $\demotions{J}[\demotionCounter] < 4 \demotionCounter+1$.
In other words, the number of claims demoted exactly~$\demotionCounter$ times is bounded above by a linear ramp with slope~4.
To prove this, we interpret~$\demotions{\roundCounter}$ as a discrete sequence and show that the effect of a round can be conservatively modeled by a convolution with a certain constant discrete sequence~$\boldsymbol{b}$.
In a sense formalized in the proof, $\demotions{1}\leq\boldsymbol{b}*\demotions{0}$, $\demotions{2}\leq\boldsymbol{b}*\boldsymbol{b}*\demotions{0}$, and so on.
The effect of multiple rounds is therefore bounded by the effect of repeated convolutions with~$\boldsymbol{b}$.
This allows us to focus the analysis on the auto-convolutions of $\boldsymbol{b}$, and it is this analysis that yields the desired result.

In the second part, after we are past $J$~rounds, we focus on distributions of claim per demotion count with a special property.
In these distributions, there is at most a single claim demoted fewer than $\demotionCounter$~times.
Even though we cannot tell the number of demotions of this single least-demoted claim, we can prove a key result:
after a \emph{constant} number of additional rounds, we reach a configuration where there is a single claim demoted fewer than $\demotionCounter+1$~times.
Since~\demotions{J} is in such a configuration for $\demotionCounter=0$, induction on~$\demotionCounter$ shows that after $J+O(\maxDemotions)$ rounds there must be a single claim left.

The complete proof goes further and bounds the constants.
It shows that the number of rounds to settlement is always less than
\begin{gather}
    13 \maxDemotions + \log_\groupSize \nSybils + 2 \sqrt{\maxDemotions \log_\groupSize \nSybils}.\label{final-theoretical-bound}
\end{gather}

\subsection{Maximum-delay strategy for adversary}\label{strategy}

From the point of view of an adversary trying to maximize the delay to finalization, Sybils with the same demotion count are fungible.
Therefore, as each round unfolds, the adversary strategic freedom is limited to groups that mix claims with different demotion counts.
We call such groups \emph{mixed}.
Recall that given a distribution of claims per demotion count, the mixed groups are completely determined by Dave's matchmaking algorithm.
For each group, one claim will win and maintain its count, whereas the count of every other claim will be decremented.
The results of all mixed groups are under the adversary's control.
Each set of choices it makes, in all their possible combinations, defines a new possible resulting distribution for the round.
Each such distribution then leads to a new partition into groups in the following round and so on, recursively, until a distribution is found with a single surviving claim.
This creates a tree structure, where each node is a distribution, connected to one child node representing each distinct distribution that could result from the adversary's choices in that round.
The root is a distribution with~\nSybils\ claims that have never been demoted.
All leaves are distributions containing a single claim.
A \emph{strategy} is a set of choices the adversary makes for each mixed group in the path from root to a leaf.
A \emph{maximum-delay strategy} is a strategy that produces the longest possible path for a given number of Sybils~\nSybils.

We have created a program that exhaustively explores the adversary's decision tree and identifies all maximum delay strategies.
Starting from the root distribution, our program dynamically generates and visits all possible child distributions.
When creating a new child distribution, the program records the parent distribution that generated it, the path length from the root, and the set of mixed-group results needed to generate the child from the parent.
Naturally, the exploration of a branch stops when it reaches a distribution with a single surviving claim.
The exploration also stops when it finds a distribution that has been previously visited but registered a longer (or equal) path length than the current visit.
If the previous path length was shorter, the node is updated with information for the current visit.
Once the exploration is complete, the program finds and prints the longest path from the initial distribution to the final one, including the mixed groups and how they broke.

We know from~\cref{asymptotic} that the maximum number of rounds is logarithmic on the number~\nSybils\ of Sybils.
Unfortunately, there can be up to~\mbox{$\maxDemotions-1$} mixed groups per round.
Therefore, each claim distribution can potentially produce~\mbox{$2^{\maxDemotions-1}$} different successors\footnote{$\groupSize>2$ can result in mixed groups with more than 2 distinct demotion counts. However, there necessarily would be fewer than $\maxDemotions-1$ of these. $2^{\maxDemotions-1}$ is the worst case for any~\groupSize.}.
There is no hope of completely exploring such a space as~\maxDemotions\ grows.
Fortunately, there is also no reason to believe that the maximum-delay strategy would change in any significant way for large~\maxDemotions.

We therefore ran our program for \mbox{$\groupSize\in\{2,3,5,9\}$}, \mbox{$\maxDemotions\in\{9,\ldots 12\}$}, and~\mbox{$\nSybils\in\{2,\ldots,4096\}$}.
In \emph{every} single case, a simple strategy produced a delay at least as large as any other:
\begin{quotation}
\emph{Preserve the claim with fewer demotions in every mixed group, except if doing so would terminate the dispute when doing otherwise wouldn't.}
\end{quotation}

This strategy makes perfect sense.
In fact, we designed Dave specifically to prevent the adversary from sacrificing highly-demoted Sybils in favor of less-demoted ones (see~\cref{matchmaking}).
Following the strategy outlined here, at some point there will be only two claims left: one that has never been demoted, and one that is about to be eliminated.
The twist at the end simply extends the dispute further since, at that point, demoting the about-to-be-eliminated claim would end the dispute.

Although it would be best to formally \emph{prove} that this is indeed a maximum-delay strategy for all \groupSize, \maxDemotions, and \nSybils, the exhaustive search offers compelling evidence that it is.

\begin{table*}[ht]
\rowcolors{6}{}{gray!30}
\centering
\caption{For a number~\nSybils\ of Sybils ranging from~$2^4$ to~$2^{32}$, an group sizes $\groupSize=2$, $\groupSize=4$, and $\groupSize=8$, the table shows the choice for~\gracePeriod\ \emph{in hours} that minimizes the worst-case dispute times~\mbox{$\maxDisputeTimeNoTc$} \emph{in days} for a no-censorship regime. Also shown are the maximum number of demotions~\maxDemotions\ used in the dispute (inversely proportional to $\gracePeriod$) and the maximum number of rounds~$\nRounds$ until settlement. For completeness, the worst-case dispute times when the adversary spends its entire censorship budget are shown as~\mbox{$\maxDisputeTimeTc = (\nRounds+\maxDemotions-1) \roundDuration$}.}
\begin{tabular}{ccS[table-format=2.1]S[table-format=2.2]ccS[table-format=2.2]cS[table-format=2.1]S[table-format=2.2]ccS[table-format=2.2]cS[table-format=2.1]S[table-format=2.2]ccS[table-format=2.2]}
\toprule
\multirow{2}{*}{\raisebox{-3pt}{\nSybils}} & \multicolumn{6}{c}{$\groupSize=2$} & \multicolumn{6}{c}{$\groupSize=4$} & \multicolumn{6}{c}{$\groupSize=8$} \\
\cmidrule(l{10pt}r{1pt}){2-7}
\cmidrule(l{10pt}r{1pt}){8-13}
\cmidrule(l{10pt}r{2pt}){14-19}
&& \multicolumn{1}{c}{\gracePeriod~(h)} & \multicolumn{1}{c}{$\maxDisputeTimeNoTc$ (d)} & \multicolumn{1}{c}{\maxDemotions} & \multicolumn{1}{c}{$\nRounds$} & \multicolumn{1}{c}{$\maxDisputeTimeTc$ (d)} &&
  \multicolumn{1}{c}{\gracePeriod~(h)} & \multicolumn{1}{c}{$\maxDisputeTimeNoTc$ (d)} & \multicolumn{1}{c}{\maxDemotions} & \multicolumn{1}{c}{$\nRounds$} & \multicolumn{1}{c}{$\maxDisputeTimeTc$ (d)} &&
  \multicolumn{1}{c}{\gracePeriod~(h)} & \multicolumn{1}{c}{$\maxDisputeTimeNoTc$ (d)} & \multicolumn{1}{c}{\maxDemotions} & \multicolumn{1}{c}{$\nRounds$} & \multicolumn{1}{c}{$\maxDisputeTimeTc$ (d)}  \\
\midrule
$2^{4}$  && 10.5 & 33.54 & 16 & 35  & 47.92 && 14.0 & 21.25 & 12 & 17  & 35.00 && 24.0 & 16.67 & 7  & 8   & 29.17 \\
$2^{8}$  && 6.2  & 43.33 & 27 & 72  & 58.98 && 5.8  & 26.61 & 29 & 47  & 42.46 && 8.0  & 21.00 & 21 & 28  & 36.00 \\
$2^{12}$ && 4.7  & 48.64 & 36 & 103 & 65.17 && 5.4  & 29.42 & 31 & 55  & 45.47 && 6.7  & 23.16 & 25 & 36  & 38.60 \\
$2^{16}$ && 3.7  & 52.86 & 45 & 134 & 70.21 && 4.4  & 31.62 & 38 & 70  & 48.34 && 6.5  & 24.87 & 26 & 40  & 40.42 \\
$2^{20}$ && 3.7  & 56.41 & 45 & 143 & 73.76 && 4.7  & 33.53 & 36 & 71  & 50.06 && 4.4  & 26.20 & 38 & 58  & 42.92 \\
$2^{24}$ && 3.3  & 59.40 & 51 & 166 & 77.29 && 3.9  & 35.17 & 43 & 86  & 52.34 && 4.7  & 27.39 & 36 & 58  & 43.92 \\
$2^{28}$ && 3.1  & 62.17 & 55 & 184 & 80.42 && 3.2  & 36.67 & 52 & 104 & 54.65 && 4.4  & 28.46 & 38 & 63  & 45.18 \\
$2^{32}$ && 2.8  & 64.76 & 61 & 207 & 83.53 && 3.4  & 38.01 & 49 & 103 & 55.73 && 3.9  & 29.44 & 43 & 72  & 46.62 \\
\bottomrule
\end{tabular} \label{tab:time-to-settle}
\end{table*}

\subsection{Worst-case delay by maximum-delay strategy}\label{tight}

\begin{figure}[t]
\includegraphics[width=\linewidth]{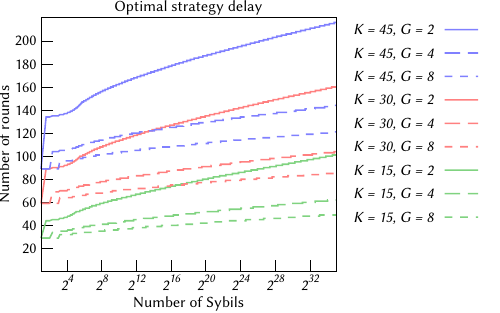}
\caption{Delay in rounds obtained by an adversary that uses the maximum-delay strategy of~\cref{strategy}.
Lines measure the maximum delay that can be achieved from~$2$ to~$2^{35}$~Sybils given a maximum demotion count~\maxDemotions\ and group size~\groupSize.
(Note the Sybils axis is in logarithmic scale.)
}
\label{30-70-plot}
\end{figure}

We can efficiently compute the delay~$\maxRoundDelay{\maxDemotions}{\groupSize}{\nSybils}$ in rounds that the maximum-delay strategy creates when claims can be demoted~\maxDemotions\ times before elimination, for any group size~\groupSize, starting from~\nSybils\ Sybils.
Better yet, since the delay is monotonic in~\nSybils, we can also quickly identify (by binary search) the next higher value of~\nSybils\ that produces a bigger delay.
Using this idea, we computed~$\maxRoundDelay{\maxDemotions}{\groupSize}{\nSybils}$ for the maximum-delay strategy of~\cref{strategy}, for all~$\groupSize\in\{2,\dots,32\}$, $\maxDemotions\in\{2\dots,80\}$, and $\nSybils\in\{2,\ldots,2^{52}\}$.
\Cref{30-70-plot} shows a selection of these results.

For each combination of maximum demotion count and group size~\mbox{$(K_j, G_j)$}, we then found the set of parameters $0 \le \alpha_j\le 13$, $0\le \beta_j \le 1$, and $0\le \gamma_j \le 2$ that minimize the quadratic error
\begin{align}
\sum_{N_i > 100} \big(\alpha_j\, K_j + \beta_j \log_{G_j} N_i + \gamma_j \sqrt{K_j \log_{G_j} N_i} - \maxRoundDelay{K_j}{G_j}{N_i}\big)^2. \nonumber
\end{align}
We excluded relatively small values of~\mbox{$\nSybils \le 100$} from the regression to skip the transient regions where the curves in figure~\ref{30-70-plot} seem to behave erratically.
The resulting fits are good.
The maximum RMS over all $(K_j, G_j)$ pairs was only 0.97 rounds, and the maximum absolute (or relative) error only 3.96 rounds (or 2.8\% for $\maxDemotions \ge 10$).

More importantly, using the maximum values for $\alpha_j,\beta_j$, and~$\gamma_j$, we get the following expression for the asymptotic bound
\begin{align}
\maxRoundDelay{\maxDemotions}{\groupSize}{\nSybils} < 2.66 \maxDemotions + \log_{\groupSize} \nSybils + 2 \sqrt{\maxDemotions \log_{\groupSize} \nSybils}. \label{final-numerical-bound}
\end{align}
This seems to indicate that the actual coefficient of \maxDemotions\ is considerably better than the theoretical bound shown in~\eqref{final-theoretical-bound}.
In fact, even taking the maximum values for $\alpha_j,\beta_j$, and~$\gamma_j$, is overly conservative.

\section{Discussion}\label{time-param}

In the previous section, we discussed the way in which the number of rounds required to resolve a dispute depends on the group size~\groupSize, the maximum demotion count~\maxDemotions, and the number of Sybils~\nSybils.
By mounting an attack on the protocol, the adversary may attempt to delay finalization by recruiting many Sybils (e.g. in a reputation attack by a rival protocol, griefing, extortion, etc.), or in exhausting the hero's resources to the point the honest claim is eliminated and a false claim is certified (to inappropriately gain access to all the value the protocol is protecting).
\Cref{liveness-analysis} addresses the former concern, and~\cref{resource} addresses the latter.
Together, they show that large attacks are not cost-effective in that they increase settlement delay only logarithmically and cannot realistically alter the result of the dispute (i.e., the TVL is safe).
By refusing to restrict participation, Dave relinquishes control over~\nSybils.
Nevertheless, it is possible to select~\groupSize\ and~\maxDemotions\ to guarantee it is secure, decentralized, and lively.

\subsection{Liveness}\label{liveness-analysis}

The worst-case total dispute time~$\maxDisputeTimeNoTc$ can be expressed as
\begin{align}
    \maxDisputeTimeNoTc &= \maxRoundDelay{\maxDemotions}{\groupSize}{\nSybils}\, \roundDuration
             = \maxRoundDelay{\maxDemotions}{\groupSize}{\nSybils}\,(2 \gracePeriod + \totalActionDuration) \\
             &= \maxRoundDelay{\maxDemotions}{\groupSize}{\nSybils}\,(2 \censorshipTimeBudget/\maxDemotions + \totalActionDuration). \label{final-numerical-time-bound}
\end{align}

Constants~\totalActionDuration, \groupSize, and~\censorshipTimeBudget\ are essentially known.
\totalActionDuration\ depends on network parameters and the time needed to generate the witness for the final step.
The speed at which a witness can be generated depends on the amount of computational power that will be expected to be available to validators.
(Bringing the time down mandates access to faster, more expensive hardware.)
We believe $\totalActionDuration = 2\ \mathrm{hours}$ to be a reasonable choice for Ethereum and relatively accessible hardware.
Likewise, increasing~\groupSize\ would improve liveness at the cost of increased resource expenditures by the hero.
Fortunately, as we will see, $\groupSize=2$ to $8$ seems adequate to guarantee liveness.
Finally, it is generally accepted that~$\censorshipTimeBudget=\mathrm{1~week}$ for Ethereum~\citep{kelvinfichter_challenge}.

The only free parameter in~\cref{final-numerical-time-bound} is the maximum number of demotions~\maxDemotions, which is connected by its definition~\mbox{$\maxDemotions=\censorshipTimeBudget/\gracePeriod$} to the grace period.
Due to resource exhaustion attacks, a longer~\gracePeriod\ is better for security.
Nevertheless, there is considerable freedom in setting this parameter.
We can therefore use~\cref{final-numerical-time-bound} to find the~\maxDemotions\ (and hence the optimal~\gracePeriod) that minimizes settlement time.
To do so, we use the worst-case delay in rounds computed in~\cref{tight}.
Since censoring transactions is expensive and technically challenging while at the same time offering little extra delay, we perform the optimization for the case in which no censorship power has been exerted.
The results are shown in~\cref{tab:time-to-settle}.
Besides the value for~\gracePeriod, the table includes the maximum dispute time~$\maxDisputeTimeNoTc$, the maximum number of demotions~\maxDemotions, and the worst-case number of rounds~$\nRounds$.
For completeness, it also includes the maximum dispute time~\mbox{$\maxDisputeTimeTc\! = (\nRounds+\maxDemotions-1)\roundDuration$} where the adversary has used its entire censorship budget.

\begin{table}[t]
\rowcolors{2}{}{gray!30}
\centering
\caption{A fixed choice for grace period~$\gracePeriod = 8\ \mathrm{hours}$ (corresponding to $\maxDemotions = 21$) and group size~$\groupSize = 4$ leads to acceptable maximum dispute times for a wide range of number~\nSybils\ of Sybils.}
\begin{tabular}{lcS[table-format=2.2]cS[table-format=2.2]}
\toprule
\nSybils && \multicolumn{1}{c}{$\maxDisputeTimeNoTc$ (d)} & $\nRounds$ & \multicolumn{1}{c}{$\maxDisputeTimeTc$ (d)} \\
\midrule
$2^{4}$  && 21.75 & 49 & 36.75 \\
$2^{8}$  && 27.00 & 56 & 42.00 \\
$2^{12}$ && 30.00 & 60 & 45.00 \\
$2^{16}$ && 33.00 & 64 & 48.00 \\
$2^{20}$ && 35.25 & 67 & 50.25 \\
$2^{24}$ && 38.25 & 71 & 53.25 \\
$2^{28}$ && 40.50 & 74 & 55.50 \\
$2^{32}$ && 42.75 & 77 & 57.75 \\
\bottomrule
\end{tabular} \label{tab:time-concrete}
\end{table}

While disputes with fewer Sybils favor a longer~\gracePeriod\ than disputes with many Sybils, dynamically picking a different value for~\gracePeriod\ based on the value of~\nSybils\ is tricky since the dispute starts \emph{before} we know the value of~\nSybils.
Fortunately, the duration~$\maxDisputeTimeNoTc$ of the dispute is not particularly sensitive to~\gracePeriod.
It is therefore possible to pick a single set of parameters that results in acceptable dispute times for a wide range of number of Sybils, striking a good balance between settlement time and resource usage.
To that end, we recommend $\gracePeriod = 8\ \mathrm{hours}$ (meaning $\maxDemotions = 21$), and $\groupSize = 4$.
\Cref{tab:time-concrete} shows the results.\footnote{Please see~\cref{proportional} (i.e., \cref{tab:time-opt-continuous-large}) for a \emph{very} promising proposal that essentially cuts these dispute times \emph{by half}.
This would allow reducing~\groupSize\ to 2 while keeping a similar liveness profile.
As we discuss in the following section, lower \groupSize\ improves security and decentralization by reducing hardware mandates, gas fees, and bonds.}

\subsection{Security and decentralization}\label{resource}

Decentralization demands that anyone is free to (and can afford to) be the hero.
Security requires that even a singleton hero can defeat an adversary backed by a nation-state.
In this section, we analyze the resources the hero needs to enter a dispute and then win it.

The first threat to security and decentralization comes from hardware mandates.
In Dave, the peak transient computational power use is independent of the number of Sybils.
After all, the hero engages simultaneously with at most a small and constant number~\mbox{$\groupSize-1$} of Sybils.
This takes only~\mbox{$\groupSize-1$} ``units of hardware''.
From the hardware perspective, Dave is therefore secure and decentralized\footnoteref{others-suck}.

The next threat comes from gas expenditure and bond requirements.
The hero's gas consumption does depend on the number of Sybils, but bonds do not---the hero makes a single claim, requiring only one bond.
It appears that security and decentralization are at odds.
Security demands that the hero's gas expenditure be low, otherwise it would be vulnerable to resource exhaustion attacks.
However, a large number of Sybils increases this expenditure.
Decentralization, on the other hand, requires that the bond price be set low so that anyone can participate.
However, this allows for the possibility of many Sybils.
In other words, expensive bonds seem to favor security over decentralization, and cheap bonds seem to favor decentralization to the detriment of security.

Fortunately, as~\cref{liveness-theorem} shows, the number of rounds grows only logarithmically with the number of Sybils.
The gas spent in the defense of a single claim is proportional to the number of rounds.
This gives the hero an exponential advantage in gas fees over the adversary.
This advantage enables the hero to afford, in terms of gas costs, a dispute against any adversary.
There is no practical amount of investment that could cause even a singleton hero's funds to run dry.
At best, the adversary can cause delays\footnote{\label{others-suck}This is in contrast with fully parallel algorithms mentioned in~\cref{related}, where a well-funded or well-equipped adversary can win. We believe that a modest cost to liveness is preferable to this security vulnerability.}.
This is perhaps the main reason we expect there will be no large-scale Sybil attacks: the adversary will never win, and delays grow logarithmically with the size of the investment that is ultimately lost.

This exponential advantage effectively decouples bond price from security---bonds can be safely set low, allowing anyone to be the hero.
Security and decentralization are not at odds, as it originally seemed.
Even a system with no bonds is secure.
Ultimately, bonds in Dave serve as a mechanism to refund the winner for resources it spent in the defense of the honest claim, and perhaps as an added incentive for acting as the hero.
Either way, a mispriced bond does not compromise the system's security.

Naturally, the bond backing the winning claim is always refunded after the dispute.
The bonds that backed \emph{losing} claims can be confiscated and awarded to the hero after the dispute is over.
(We assume a singleton hero for simplicity, but see \cref{aa} for future work on reimbursing or rewarding a collective hero.)
The accounting is very favorable.
After all, the hero's expenses increase only logarithmically with the increasing number of confiscated bonds.
The worse-case scenario is really for $\nSybils = 2$ (hero versus one Sybil).

Consider this worst-case scenario under the parameters shown in~\cref{tab:time-concrete}.
We estimate that, at current Ethereum costs, the gas expenses for a single match are under~\mbox{$C_m = \tfrac{1}{20}\ \mathrm{ether}$}.
The hero spends at most~\mbox{$(\groupSize-1)\,C_m$} per round.
Since~\mbox{$\maxDemotions=21$} and $\nSybils = 2$, the hero participates in at most $\nRounds=21$ rounds.
It spends a maximum of~\mbox{$(\groupSize-1)\, C_m\, \nRounds \approx 3\ \mathrm{ether}$} during the entire dispute.
We therefore set the bonds to 3~ether.

Now consider a scenario where the adversary commands a mob with one million coordinated Sybils.
With each bond costing 3~ether, the adversary ultimately loses~\emph{3~million~ether} in confiscated bonds.
The hero, on the other hand, only needs to allocate 10~ether, which it will recover after the dispute: \mbox{$(\groupSize-1)\, C_m\, \nRounds \approx 7\ \mathrm{ether}$} in expenses and the 3~ether bond.
This level of discrepancy guarantees the hero's triumph over extremely well-funded adversaries.
Further expenditures by the adversary do not pay off since both the hero's expenses and resulting delay grow logarithmically.
\Cref{tab:comparison} shows a comparison between competing algorithms under an attack where the adversary invests the same resources.
Note how Dave strikes a favorable balance between bonds, expenses, and delay relative to competing methods.
Although this is certainly a very expensive attack, BoLD currently protects almost twice this amount (see L2BEAT).
It is therefore conceivable if it has an appreciable chance of succeeding.

\begin{table}[t]
\centering
\caption{Comparison of cost to hero for a 3 million ether adversary attack on different algorithms. Bond and expenses are refunded to hero.}
\rowcolors{2}{}{gray!30}
\begin{threeparttable}
\begin{tabular}{lccc}
\toprule
& Bond & Expenses & Delay \\
\midrule
Dave\tnote{$\dagger$}  & 3 ether          & 7 ether            &  5 weeks \\
PRT-1L                 & 1 ether          & 1 ether            & 22 weeks \\
BoLD-1L\tnote{*}       & \num{1000} ether & \num{3000} ether   &  1 week  \\
BoLD-3L\tnote{*}       & \num{3600} ether & \num{461539} ether &  1 week  \\
\bottomrule
\end{tabular}
\begin{tablenotes}
\item[$\dagger$] Using~$\groupSize=4$ and~$\maxDemotions=21$.
\item[*] Data from~\citet{Arbitrum-Economics} and \citet{fraud-proof-war}.
\end{tablenotes}
\end{threeparttable}\label{tab:comparison}
\end{table}

\section{Conclusion}

In this paper we presented a new fraud-proof algorithm that offers a unique balance between decentralization, security, and liveness.
The barriers to entry, both in terms of bonds and in terms of hardware mandates, are constant and very low.
This allows anyone to participate (decentralization).
The quantity of resources necessary to mount an attack is exponential both on the delay it manages to cause and on the resources needed to overcome it.
This makes it impractical to defeat the honest claim (security).
Finally, a new strategy for amortizing censorship over the entire dispute enables punishing unresponsiveness without risking security or introducing large delays.
In practice, no dispute will take longer than 2--5 weeks to complete (liveness).

In the following sections, we discuss a variety of ideas with the potential to improve these results even further.
Chief among them, a variation of the algorithm that leads to reduced round durations and, consequently, significantly improved liveness.

\subsection{Reducing the duration of rounds}\label{proportional}

We propose a variation on our algorithm that can reduce the round duration from~$\roundDuration=\totalActionDuration+2\gracePeriod$ to $\roundDuration'=\totalActionDuration+\gracePeriod$.
Since~$\gracePeriod$ is considerably longer than~$\totalActionDuration$, this effectively halves the dispute time.
\Cref{tab:time-opt-continuous-large} shows the improvements in liveness using the same~\gracePeriod\ as~\cref{tab:time-concrete}.
The method makes~$\groupSize=2$ attractive and~$\groupSize=4$ even better.

Rather than counting demotions, we will now precisely measure the amount of time~\censorshipTimeAccumulated\ during which a claim must have been censored, if it is honest and properly defended.
Initially, we set~\mbox{$\censorshipTimeAccumulated=0$} for each claim.
Groups are formed in increasing order of~\censorshipTimeAccumulated.
Each round lasts exactly~$\roundDuration'=\totalActionDuration+\gracePeriod$.
During each match, for each claim, we add up the amount of time~\actionTimeSpent\ elapsed while it was its turn to act.
After the round is over, for each claim, we keep the maximum~\actionTimeSpent\ over all matches it played within its group.
We then add to each claim's~\censorshipTimeAccumulated\ the corresponding~$\max(0, \actionTimeSpent-\tfrac{1}{2}\totalActionDuration)$.
By assumption, no claim can be censored longer than~\censorshipTimeBudget.
As soon as a claim has been ``censored'' for~$\censorshipTimeAccumulated > \censorshipTimeBudget$, we know it cannot be honest or properly defended, so it is fair to eliminate it from the dispute.
Until there is a single claim left, we reform groups and continue the process.

\begin{table}[t]
\rowcolors{3}{gray!30}{}
\centering
\caption{Effect of reduced round duration when $\gracePeriod = 8\ \mathrm{hours}$.}
\begin{tabular}{ccS[table-format=2.2]cS[table-format=2.2]cS[table-format=2.2]cS[table-format=2.2]}
\toprule
\multirow{2}{*}{\raisebox{-3pt}{\nSybils}} & \multicolumn{4}{c}{$\groupSize=2$} & \multicolumn{4}{c}{$\groupSize=4$} \\
\cmidrule(l{10pt}r{1pt}){2-5}
\cmidrule(l{10pt}r{1pt}){6-9}
&& \multicolumn{1}{c}{$\maxDisputeTimeNoTc$ (d)} & $\nRounds$ & \multicolumn{1}{c}{$\maxDisputeTimeTc$ (d)}
&& \multicolumn{1}{c}{$\maxDisputeTimeNoTc$ (d)} & $\nRounds$ & \multicolumn{1}{c}{$\maxDisputeTimeTc$ (d)} \\
\midrule
$2^{4}$  && 18.75 & 45  & 27.08 &&  12.08 & 29 & 20.42 \\
$2^{8}$  && 24.58 & 59  & 32.92 &&  15.00 & 36 & 23.33 \\
$2^{12}$ && 28.33 & 68  & 36.67 &&  16.67 & 40 & 25.00 \\
$2^{16}$ && 31.25 & 75  & 39.58 &&  18.33 & 44 & 26.67 \\
$2^{20}$ && 34.17 & 82  & 42.50 &&  19.58 & 47 & 27.92 \\
$2^{24}$ && 37.08 & 89  & 45.42 &&  21.25 & 51 & 29.58 \\
$2^{28}$ && 39.58 & 95  & 47.92 &&  22.50 & 54 & 30.83 \\
$2^{32}$ && 42.08 & 101 & 50.42 &&  23.75 & 57 & 32.08 \\
\bottomrule
\end{tabular}\label{tab:time-opt-continuous-large}
\end{table}

When there is no censorship and no deliberate unresponsiveness, \mbox{$\actionTimeSpent \le \tfrac{1}{2}\totalActionDuration$} and~\censorshipTimeAccumulated\ is left unchanged.
A properly defended honest claim's~\censorshipTimeAccumulated\ is incremented by exactly as much censorship as was imposed on it.
Matches between dishonest claims either complete before the round is over, or not.
If they complete, one of the claims will have been eliminated from the dispute.
If they do not, their~\censorshipTimeAccumulated\ will have been incremented in proportion to how much they ran down their own clocks on purpose, from a total of~\gracePeriod.

There is a strong analogy with the algorithm presented earlier.
The demotion count of a claim is nothing but a discrete version of~\censorshipTimeAccumulated, in units of~\gracePeriod.
It also measures the amount of censorship that a claim must have suffered, if it is honest and properly defended.
By assumption, no claim can be censored for more than $\censorshipTimeBudget=\maxDemotions\,\gracePeriod$.
Therefore, a claim demoted more than~\maxDemotions~times cannot be honest or properly defended.

In the discrete variant, we have seen that the adversary's optimal delay strategy is to safeguard the demotion count of its least demoted Sybils.
The analogous strategy in the continuous variant would be to safeguard the~\censorshipTimeAccumulated\ of the Sybils with smallest~\censorshipTimeAccumulated.
Under the assumption that this is indeed optimal, it is easy to see that the worst-case dispute would progress in the same way.
The only difference is the lower round duration of the continuous variant.
Hence the numbers shown in~\cref{tab:time-opt-continuous-large}.

Finding an asymptotic bound for the continuous variant seems harder than in the discrete case.
Even running a brute-force search for the optimal delay strategy seems challenging.
Nevertheless, we conjecture that the analogy with the discrete variant holds.
Given how exciting the results are, we strongly recommend this as a priority for future work.

\subsection{Helping the hero}\label{aa}

In this article we have considered the situation in which a single entity acts as the hero.
By its nature, computation hashes allow multiple parties to collaborate in the defense of the honest claim, without trusting each other.
We should design and put in place a system of incentives that encourages wider participation.

One idea is to distribute the bonds lost by all Sybils to those that helped defend the honest claim.
A first step in this direction could be the creation of a token for each claim.
These tokens would be distributed whenever someone performs a successful action in defense of its corresponding claim.
Once the dispute is over, the tokens associated to the honest claim could be redeemed for a proportional fraction of the locked bonds.
Although a naive implementation of this idea would be vulnerable to Dark Forest attacks (see~\cite{dark-forest}),
ZK commitments could be used to side-step this problem.

Another improvement is to use locked bonds to sponsor all bisection and step actions.
As a consequence, the hero's expenses become constant, further improving security.
This could be accomplished with \emph{account abstraction}.
Once a claim was made and the corresponding bond is locked, all further actions would be free and could be submitted from ad-hoc wallets.

\subsection{Additional future work}

It is possible for PRT to complete faster than Dave when the number of Sybils is very small (at least when we do not use the improvements of~\cref{proportional}).
An implementation might want to combine both algorithms, for example by running the two in parallel and using the earliest result.

Although we have considered a constant group size~\groupSize, it would be interesting to investigate the improvements to liveness achievable by setting, for example, $G = \log N$.
This would guarantee groups are still relatively small.
For larger group sizes, it might be important to adopt a different clock designed for that purpose, such as the one proposed by BoLD.

\appendix

\section{Proof of asymptotic bound}\label{liveness-proof}

We now formally prove
\begin{reptheorem}{liveness-theorem}
When groups have at most~\groupSize~claims, and a claim can be demoted \maxDemotions~times before being eliminated, Dave settles a dispute involving \nSybils~claims in $O\big(\maxDemotions + \log_\groupSize \nSybils + \sqrt{\maxDemotions \log_\groupSize\nSybils}\,\big)$ rounds.
\end{reptheorem}

Let~$\demotions{\roundCounter}[\demotionCounter]$ denote the number of claims demoted exactly~$\demotionCounter$ times after~\mbox{$\roundCounter = 0, 1, \dots$}~rounds.
It will be more convenient to manipulate the distribution of demotions~\demotions{\roundCounter}\ using operations over discrete sequences.
To that end, we define scalar multiplication and discrete convolution between two discrete sequences~$\boldsymbol{x}$ and~$\boldsymbol{y}$ in the usual fashion:
\begin{alignat}{3}
\boldsymbol{y} &=\alpha \boldsymbol{x},\quad && \text{if }\boldsymbol{y}[k]=\alpha\boldsymbol{x}[k]\text{ for all $k\in\mathbb{Z}$, and}\\
\boldsymbol{z} &= \boldsymbol{x}*\boldsymbol{y},\quad && \text{if }\boldsymbol{z}[k] = \sum_{\mathclap{j=-\infty}}^\infty \boldsymbol{x}[j] \boldsymbol{y}[k-j]\text{ for all $k\in \mathbb{Z}$.} \\
\intertext{We also define point-wise comparisons on a half-line, for $i\in\mathbb{Z}$,}
\boldsymbol{x} &\lel{i} \boldsymbol{y},\quad  && \text{if }\boldsymbol{x}[k] \leq \boldsymbol{y}[k]\text{ for all $k \ge i$ ($i=0$ if omitted)}.
\end{alignat}
In addition, we define the discrete sequences \emph{unit impulse}~$\boldsymbol{\delta}_i$, \emph{unit step}~$\boldsymbol{u}_i$, \emph{unit linear ramp}~$\boldsymbol{r}_i$, and the \emph{binomial kernel}~$\boldsymbol{b}$ so that
\begin{align}
\boldsymbol{\delta}_i[k] &= \begin{cases}
        1, & k = i, \\
        0, & \text{otherwise,} \\
    \end{cases}\\
\boldsymbol{u}_i[k] &=
    \begin{cases}
        1, & k \ge i, \\
        0, &\text{otherwise,}
    \end{cases} \\
\boldsymbol{r}_i[k] &= \begin{cases}
        k-i+1, & k\ge i,\\
        0, & \text{otherwise, and}
    \end{cases}\\
\boldsymbol{b}[k] &=
    \begin{cases}
       q, & k = 0, \\
       p, & k = 1, \\
       0, & \text{otherwise.} \\
    \end{cases}
\end{align}
Finally, abusing notation, we denote repeated auto-convolution by
\begin{align}
\boldsymbol{b}^{i} &= \begin{cases}
        \boldsymbol{\delta}_0, & i = 0, \\
        \boldsymbol{b}*\boldsymbol{b}^{i-1}, & i>0.
    \end{cases}
\end{align}

Initially, we have~\mbox{$\demotions{\,0}[0] = \nSybils$}, and~\mbox{$\demotions{\,0}[\demotionCounter] = 0$} for~\mbox{$\demotionCounter \neq 0$}.
Although all claims start out with zero demotions, as we will demonstrate, the distribution of claims into~\demotions{\roundCounter}[\demotionCounter]\ quickly ``flattens'' and pushes claims towards elimination.

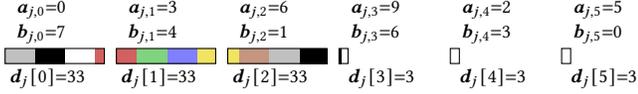
\begin{figure}
  \centering
  \begin{tikzpicture}[scale=0.04]
    \fill[color=a]
    (37 * 0 + 0, 0) rectangle (37 * 0 + 10, 5);
    \fill[color=b]
    (37 * 0 + 10, 0) rectangle (37 * 0 + 20, 5);
    \fill[color=c]
    (37 * 0 + 20, 0) rectangle (37 * 0 + 30, 5);
    \fill[color=d]
    (37 * 0 + 30, 0) rectangle (37 * 0 + 33, 5);
    \fill[color=d]
    (37 * 1 + 0, 0) rectangle (37 * 1 + 7, 5);
    \fill[color=e]
    (37 * 1 + 7, 0) rectangle (37 * 1 + 17, 5);
    \fill[color=f]
    (37 * 1 + 17, 0) rectangle (37 * 1 + 27, 5);
    \fill[color=g]
    (37 * 1 + 27, 0) rectangle (37 * 1 + 33, 5);
    \fill[color=g]
    (37 * 2 + 0, 0) rectangle (37 * 2 + 4, 5);
    \fill[color=h]
    (37 * 2 + 4, 0) rectangle (37 * 2 + 14, 5);
    \fill[color=i]
    (37 * 2 + 14, 0) rectangle (37 * 2 + 24, 5);
    \fill[color=j]
    (37 * 2 + 24, 0) rectangle (37 * 2 + 33, 5);
    \fill[color=j]
    (37 * 3 + 0, 0) rectangle (37 * 3 + 1, 5);
    \fill[color=k]
    (37 * 3 + 1, 0) rectangle (37 * 3 + 3, 5);
    \fill[color=k]
    (37 * 4 + 0, 0) rectangle (37 * 4 + 3, 5);
    \fill[color=k]
    (37 * 5 + 0, 0) rectangle (37 * 5 + 3, 5);
    \foreach \i in {0, 1, 2} {
      \draw (37 * \i + 0, 0) rectangle (37 * \i + 33, 5);
      \node at (37 * \i + 14, -5)
      {$\scriptstyle \demotions{\roundCounter}[\i] = 33$};
    }
    \foreach \i in {3, 4, 5} {
      \draw (37 * \i + 0, 0) rectangle (37 * \i + 3, 5);
      \node at (37 * \i + 14, -5)
      {$\scriptstyle \demotions{\roundCounter}[\i] = 3$};
    }
    \node at (0 * 37 + 12, 18)
    {$\scriptstyle \borrowFirst{\roundCounter}{0} = 0$};
    \node at (0 * 37 + 12, 10)
    {$\scriptstyle \borrowLast{\roundCounter}{0} = 7$};
    \node at (1 * 37 + 12, 18)
    {$\scriptstyle \borrowFirst{\roundCounter}{1} = 3$};
    \node at (1 * 37 + 12, 10)
    {$\scriptstyle \borrowLast{\roundCounter}{1} = 4$};
    \node at (2 * 37 + 12, 18)
    {$\scriptstyle \borrowFirst{\roundCounter}{2} = 6$};
    \node at (2 * 37 + 12, 10)
    {$\scriptstyle \borrowLast{\roundCounter}{2} = 1$};
    \node at (3 * 37 + 12, 18)
    {$\scriptstyle \borrowFirst{\roundCounter}{3} = 9$};
    \node at (3 * 37 + 12, 10)
    {$\scriptstyle \borrowLast{\roundCounter}{3} = 6$};
    \node at (4 * 37 + 12, 18)
    {$\scriptstyle \borrowFirst{\roundCounter}{4} = 2$};
    \node at (4 * 37 + 12, 10)
    {$\scriptstyle \borrowLast{\roundCounter}{4} = 3$};
    \node at (5 * 37 + 12, 18)
    {$\scriptstyle \borrowFirst{\roundCounter}{5} = 5$};
    \node at (5 * 37 + 12, 10)
    {$\scriptstyle \borrowLast{\roundCounter}{5} = 0$};
  \end{tikzpicture}
  \caption{Example of the groups formed when $\maxDemotions=5$ and $\groupSize=10$.
    In the example, there are $33$~claims each with 0, 1, or 2~demotions, and $3$ claims each with 3, 4, or 5~demotions.
    Different groups are shown in different colors.
    The values borrowed from left ($\borrowFirst{\roundCounter}{\demotionCounter}$) and right ($\borrowLast{\roundCounter}{\demotionCounter}$) have are also shown.}
  \label{f:groups}
\end{figure}

Dave forms groups in increasing order of demotion count.
In particular, all groups that contain at least one claim from~$\demotions{\roundCounter}[\demotionCounter]$ are formed before any group is formed that contains only claims from~$\demotions{\roundCounter}[\demotionCounter+1]$ and higher demotion counts.
To help us cope with mixed groups, which include claims with distinct demotion counts, we define quantities~$\borrowFirst{\roundCounter}{\demotionCounter}$ and~$\borrowLast{\roundCounter}{\demotionCounter}$.
For $\demotionCounter > 0$, $\borrowFirst{\roundCounter}{\demotionCounter}$ counts the number of claims that the first group to be formed containing a claim from~\demotions{\roundCounter}[\demotionCounter]\ uses from those left over by~$\demotions{\roundCounter}[\demotionCounter-1]$ or still lower demotion counts $\demotionCounter$.
Likewise, $\borrowLast{\roundCounter}{\demotionCounter}$ counts the number or claims that the last group with a claim from~\demotions{\roundCounter}[\demotionCounter]\ uses from~$\demotions{\roundCounter}[\demotionCounter+1]$ or still higher demotion counts~$\demotionCounter$.
(See \cref{f:groups} for an example.)
With these definitions in hand, we can write
\begin{gather}
\hspace*{-1ex}\text{\big(\# of groups with a claim in $\demotions{\roundCounter}[\demotionCounter]$\big)} = \tfrac{1}{\groupSize}\big(\demotions{\roundCounter}[\demotionCounter] + \borrowFirst{\roundCounter}{\demotionCounter} + \borrowLast{\roundCounter}{\demotionCounter}\big) .
\end{gather}

The number~$\demotions{\roundCounter}[\demotionCounter]$ of claims with~\demotionCounter\ demotions after~\roundCounter\ rounds will be a combination of those in~$\demotions{\roundCounter-1}[\demotionCounter]$ that won round~$\roundCounter-1$ (at most one per group) and those in~$\demotions{\roundCounter-1}[\demotionCounter-1]$ that lost round~$\roundCounter-1$.
Assuming $\borrowFirst{\roundCounter-1}{\demotionCounter-1}\neq 0$ and $\borrowLast{\roundCounter-1}{\demotionCounter-1} \neq 0$, and setting $q=\tfrac{1}{\groupSize}$ and $p=1-q$, we obtain
\begin{align}
\begin{split}
\demotions{\roundCounter}[\demotionCounter]
    & \leq p\, \big(\demotions{\roundCounter-1}[\demotionCounter-1] - ({\scriptstyle \groupSize} - \borrowFirst{\roundCounter-1}{\demotionCounter-1}) - ({\scriptstyle \groupSize} - \borrowLast{\roundCounter-1}{\demotionCounter-1})\big)\!\!\! \\
     & \quad + ({\scriptstyle \groupSize} - \borrowFirst{\roundCounter-1}{\demotionCounter-1}) + ({\scriptstyle \groupSize} - \borrowLast{\roundCounter-1}{\demotionCounter-1}) \\
     & \quad + q\, \big(\demotions{\roundCounter-1}[\demotionCounter] + \borrowFirst{\roundCounter-1}{\demotionCounter} + \borrowLast{\roundCounter-1}{\demotionCounter}\big).
\end{split}
\label[ineq]{initial-recurrence-with-a-b}
\end{align}
The first row in~\cref{initial-recurrence-with-a-b} counts the number of groups formed by members of~$\demotions{\roundCounter-1}[\demotionCounter-1]$ exclusively, where each group sends \mbox{$\groupSize-1$}~losing claims to $\demotions{\roundCounter}[\demotionCounter]$.
The second row sends to $\demotions{\roundCounter}[\demotionCounter]$ all claims from $\demotions{\roundCounter-1}[\demotionCounter-1]$ that are in the first and last group: those that respectively include claims from lower and higher demotion counts~$\demotionCounter$.
The third row counts the claims from~$\demotions{\roundCounter-1}[\demotionCounter]$ that won their group and remain in $\demotions{\roundCounter}[\demotionCounter]$.
The maximum is reached when \mbox{$\borrowFirst{\roundCounter-1}{\demotionCounter-1} = \borrowLast{\roundCounter-1}{\demotionCounter-1} = 1$} and \mbox{$\borrowFirst{\roundCounter-1}{\demotionCounter} = \borrowLast{\roundCounter-1}{\demotionCounter} = \groupSize-1$}, so that
\begin{align}
\demotions{\roundCounter}[\demotionCounter]
    & \leq p\, \demotions{\roundCounter-1}[\demotionCounter-1]
     + q\, \demotions{\roundCounter-1}[\demotionCounter]
     + 4 p.
\label[ineq]{scalar-round}
\end{align}
(The cases where $\borrowFirst{\roundCounter-1}{\demotionCounter-1} = 0$ or $\borrowLast{\roundCounter-1}{\demotionCounter-1} = 0$ lead to smaller bounds.)
Since~$\demotions{\roundCounter}[k] = 0$ for~$k < 0$, we have a tighter bound for~\demotions{\roundCounter}[0]:
\begin{align}
\demotions{\roundCounter}[0]
    &\leq q\, \demotions{\roundCounter-1}[0] + p.
\label[ineq]{special-scalar-round}
\end{align}

Using our notation, we can rewrite~\cref{scalar-round,special-scalar-round} as
\begin{align}
    \demotions{\roundCounter}
         & \leq \boldsymbol{b}*\demotions{\roundCounter-1} + \boldsymbol{u}'_0,\quad\text{with $\boldsymbol{u}'_i= p\,(\boldsymbol{u}_i + 3 \boldsymbol{u}_{i+1})$} \\
\intertext{which we can keep iterating}
         & \leq \boldsymbol{b}*\big(\boldsymbol{b}*\demotions{\roundCounter-2} + \boldsymbol{u}'_0\big) + \boldsymbol{u}'_0 \\
         & \leq \boldsymbol{b}*\Big(\boldsymbol{b}*\big(\boldsymbol{b}*\demotions{\roundCounter-3} + \boldsymbol{u}'_0\big) + \boldsymbol{u}'_0\Big) + \boldsymbol{u}'_0\qquad\ldots
\end{align}
until we reach~\mbox{$\demotions{\,0}=\nSybils \boldsymbol{\delta}_0$}.
Linearity, associativity, distributivity, and commutativity apply amply, so that:
\begin{align}
    \demotions{\roundCounter}
      &\leq \nSybils \boldsymbol{b}^{\roundCounter} + \left(\sum_{i=0}^{\roundCounter-1} \boldsymbol{b}^{i}\right)*\boldsymbol{u}'_0.\label[ineq]{iterated-with-summation}
\end{align}

The summation can be calculated much like a geometric series.
To that end, let
\begin{align}
\boldsymbol{s} = \sum_{i=0}^{\roundCounter-1} \boldsymbol{b}^{i}
\end{align}
Then,
\begin{align}
\boldsymbol{s}*(\boldsymbol{\delta}_0-\boldsymbol{b})= \boldsymbol{\delta}_0 - \boldsymbol{b}^\roundCounter
\end{align}
so that
\begin{align}
\boldsymbol{s} = (\boldsymbol{\delta}_0 - \boldsymbol{b}^\roundCounter)*(\boldsymbol{\delta}_0-\boldsymbol{b})^{-1}.
\end{align}

Substituting into~\cref{iterated-with-summation}, we obtain
\begin{align}
    \demotions{\roundCounter}
      \leq \nSybils \boldsymbol{b}^{\roundCounter} + (\boldsymbol{\delta}_0 - \boldsymbol{b}^\roundCounter)*(\boldsymbol{\delta}_0-\boldsymbol{b})^{-1}*\boldsymbol{u}'_0.\label[ineq]{iterated-with-inverse}
\end{align}

To understand the inverse~\mbox{$(\boldsymbol{\delta}_0-\boldsymbol{b})^{-1}$}, consider~$(\boldsymbol{\delta}_0-\boldsymbol{b})$ itself first.
If~$\boldsymbol{y}=(\boldsymbol{\delta}_0-\boldsymbol{b})*\boldsymbol{x}$, then
\begin{align}
    \boldsymbol{y}[k] &= p \big(\boldsymbol{x}[k] - \boldsymbol{x}[k-1]\big).\label{to-invert}
\end{align}
The inverse of~\cref{to-invert} is simply the causal recursive filter
\begin{align}
    \boldsymbol{x}[k] &= \tfrac{1}{p} \boldsymbol{y}[k] +\boldsymbol{x}[k-1],
\end{align}
so that
\begin{align}
  (\boldsymbol{\delta}_0-\boldsymbol{b})^{-1}*\boldsymbol{u}_i &=
  \tfrac{1}{p}\boldsymbol{r}_i.
\end{align}

This simplifies~\cref{iterated-with-inverse} to
\begin{align}
    \demotions{\roundCounter} &\leq
        \nSybils \boldsymbol{b}^{\roundCounter} + (\boldsymbol{\delta}_0 - \boldsymbol{b}^\roundCounter) *\boldsymbol{r}'_0,\quad\text{with
    $\boldsymbol{r}'_i = \boldsymbol{r}_i + 3 \boldsymbol{r}_{i+1}$}
\end{align}
and since $\boldsymbol{b}^\roundCounter[k] \ge 0$ and $\boldsymbol{r}'[k] \ge 0$ for all $k$, simplifies it further to
\begin{align}
    \demotions{\roundCounter}
      &\leq \nSybils \boldsymbol{b}^{\roundCounter} + \boldsymbol{r}'_0. \label[ineq]{iterated-with-power}
\end{align}

We now seek~$J$ such that when $j \ge J$ we have
\begin{align}
  \boldsymbol{b}^{\roundCounter}[\demotionCounter] &< \frac{1}{N},\quad\text{for $0 \leq \demotionCounter < \maxDemotions$},
\end{align}
because then, after~$J$~rounds, rounding down to integers,
\begin{align}
    \demotions{\roundCounter} &\leq \boldsymbol{r}'_0. \label[ineq]{iterated}
\end{align}

To find such~$J$, we analyze the repeated auto-convolution~$\boldsymbol{b}^{\roundCounter}$.
This is easiest in the z-domain, where convolutions become products.
The Z-transform~$\mathcal{B}$ of~$\boldsymbol{b}$ is
\begin{gather}
  \mathcal{B}(z) = \sum_{\mathclap{k=-\infty}}^\infty \boldsymbol{b}[k]\, z^{-k}\! = q + p\, z^{-1}.
\end{gather}

The transform of the repeated auto-convolution is therefore
\begin{gather}
  \mathcal{B}^\roundCounter(z) = \big(q + p\, z^{-1}\big)^\roundCounter
  = \sum_{k=0}^\roundCounter \tbinom{\roundCounter}{k}\; p^k q^{\roundCounter-k} z^{-k},
\end{gather}
and inverting it we obtain
\begin{gather}
  \boldsymbol{b}^{\roundCounter}[k] =
  \begin{cases}
    \tbinom{\roundCounter}{k}\; p^k q^{\roundCounter-k}, & k\in\{0,\ldots,\roundCounter\}, \\
    0, & \text{otherwise.}
  \end{cases} \label{auto-convolution}
\end{gather}

The equation
\begin{align}
\binom{\roundCounter}{k}\; p^k q^{\roundCounter-k} &= \frac{1}{\nSybils}
\end{align}
is unwieldy.
Fortunately, noting that
\begin{align}
\binom{\roundCounter}{k} &< \left(\frac{\roundCounter e}{k}\right)^{k},\quad\text{for $1 \le k \le \roundCounter$},\label[ineq]{binomial-bound}
\end{align}
we can instead solve
\begin{align}
\left(\frac{\roundCounter e}{k}\right)^{k} p^k q^{\roundCounter-k} &= \frac{1}{\nSybils}.
\end{align}

The two real roots are at
\begin{gather}
j_0 = \frac{\maxDemotions\, W_0(\alpha)}{\log q}\quad\text{and}\quad j_{-1} = \frac{\maxDemotions\, W_{-1}(\alpha)}{\log q},\\
\intertext{where $W_0$ and $W_{-1}$ are branches of the Lambert~$W$ function, and}
\alpha = \frac{q \log q}{e p \nSybils^{1/\maxDemotions}}.
\end{gather}

The root we want is $j_{-1}$.
We can use the following bound, due to~\citet{chatzigeorgiou2013bounds}, to replace $W$ with logarithms:
\begin{align}
W_{-1}(x) > \log (-x) -\sqrt{-2\big(1+\log(-x)\big)},\quad\text{for $-\tfrac{1}{e} < x < 0$},
\end{align}
to reach
\begin{align}
j_{-1} < \frac{\maxDemotions \log (-\alpha )}{\log q}-\frac{\sqrt{2} \maxDemotions \sqrt{-\log (-e \alpha )}}{\log q}.
\end{align}
Looking closely at the two terms,
\begin{align}
 \frac{\maxDemotions \log (-\alpha )}{\log q} &=
    \log_\groupSize \nSybils + \maxDemotions \log_\groupSize \tfrac{e (\groupSize-1)}{\log \groupSize},\quad\text{and} \\
-\frac{\sqrt{2} \maxDemotions \sqrt{-\log (-e \alpha )}}{\log q} &\le
    \sqrt{2 \maxDemotions \frac{\log_\groupSize \nSybils}{\log \groupSize}} + \maxDemotions \sqrt{2 \frac{\log_\groupSize \left(\frac{\groupSize-1}{\log \groupSize}\right)}{\log \groupSize}},
\end{align}
so we finally obtain
\begin{align}
\demotions{J} &\le \boldsymbol{r}'_0,\quad\text{with} \label[ineq]{induction-base} \\
J &= \lceil j_{-1}\rceil < 4 \maxDemotions + \log_\groupSize \nSybils + 2 \sqrt{\maxDemotions \log_\groupSize \nSybils}. \label{big-J-bound}
\end{align}

Now define proposition~$\invariant{J_\demotionCounter}{\demotionCounter}$ to mean \emph{there is at most one claim that has been demoted~\demotionCounter\ or fewer times after~$J_\demotionCounter$~rounds}.
We will complete the proof using induction on~\demotionCounter.
We will first find~$\invariantInc{0}$ so that~$\invariant{J_0}{0}$ for~$J_0 = J+\invariantInc{0}$.
Then, assuming~\invariant{J_{\demotionCounter-1}}{\demotionCounter-1}\ holds, we will show that we can find~$\invariantInc{\demotionCounter-1}$ so that~\invariant{J_\demotionCounter}{\demotionCounter}\ holds for $J_\demotionCounter=J_{\demotionCounter-1}+\invariantInc{\demotionCounter}$.
When we reach $J_{\maxDemotions-1}$, there will be at most one claim left and we are done.
At that point, the number of rounds elapsed will be at most
\begin{align}
J_{\maxDemotions-1} = J + \sum_{\demotionCounter=0}^{\maxDemotions-1} \invariantInc{\demotionCounter}.\label{last-round-summation}
\end{align}

From~\cref{induction-base}, we see that~\invariant{J_0}{0}\ holds with $J_0 = J+\invariantInc{0}$ and $\invariantInc{0}=0$.
This covers the induction base.

We now move to the induction step.
From now on, we will call \emph{leader} the claim with fewest demotions (in case of draw, one can be chosen arbitrarily).
We also define~\leastDemotions{\roundCounter}\ as demotion count of the leader after \roundCounter~rounds.
Assume~\invariant{J_{\demotionCounter-1}}{\demotionCounter-1} holds and let $J_{\demotionCounter-1} \le \roundCounter < J_{\demotionCounter}$ (whatever $J_{\demotionCounter}$ might be).
When $\leastDemotions{\roundCounter} = \demotionCounter$, the situation is analogous to~\cref{special-scalar-round} and
\begin{align}
\demotions{\roundCounter+1}[\demotionCounter] &\leq q\, \demotions{\roundCounter}[\demotionCounter] + p. \label[ineq]{highest-rank-past}
\end{align}

When $\leastDemotions{\roundCounter} < \demotionCounter$, \cref{initial-recurrence-with-a-b} reduces to
\begin{align}
\demotions{\roundCounter+1}[\demotionCounter]
    &\leq 1 + q\, \demotions{\roundCounter}[\demotionCounter] + 1, \label[ineq]{highest-rank-smaller}
\end{align}
because there is a single claim that could be demoted into $\demotions{\roundCounter+1}[\demotionCounter]$ \big(i.e., first two rows in~\eqref{initial-recurrence-with-a-b} can't exceed 1\big) and, likewise, a single group with claims in $\demotions{\roundCounter}[\demotionCounter]$ borrows that single claim with fewer demotions (i.e., $\borrowFirst{\roundCounter}{\demotionCounter}=1$ and $\borrowLast{\roundCounter}{\demotionCounter}\leq \groupSize-1$).

This bound for $\leastDemotions{\roundCounter} < k$ can be tightened for all but a single round:\\
If $\leastDemotions{\roundCounter+1} = \leastDemotions{\roundCounter}$, then the leader claim is not in~$\demotions{\roundCounter}[\demotionCounter]$ and has won its group.
This means it was not demoted to $\demotions{\roundCounter}[\demotionCounter]$ \big(one fewer than~\eqref{highest-rank-smaller}\big), and all other members in its group ended up being demoted more than \demotionCounter~times \big(another fewer than~\eqref{highest-rank-smaller}\big).
Therefore, when $\leastDemotions{\roundCounter+1} = \leastDemotions{\roundCounter}$,
\begin{align}
\demotions{\roundCounter+1}[\demotionCounter] &\leq q\, \demotions{\roundCounter}[\demotionCounter].\label[ineq]{highest-rank-fixed}
\end{align}
If $\leastDemotions{\roundCounter+1} = \leastDemotions{\roundCounter}+1 < k $, which will happen exactly $\leastDemotions{J_{\demotionCounter}}-\leastDemotions{J_{\demotionCounter-1}}$ times as $\roundCounter$ ranges from $J_{\demotionCounter-1}$ to $J_{\demotionCounter}-1$, the leader claim, which is not in~$\demotions{\roundCounter}[\demotionCounter]$, lost its group but has not yet been demoted into $\demotions{\roundCounter}[\demotionCounter]$.
This adds one back to the previous case:
\begin{align}
\demotions{\roundCounter+1}[\demotionCounter] &\leq q\, \demotions{\roundCounter}[\demotionCounter] + 1.\label[ineq]{highest-rank-moving}
\end{align}
It is only when $\leastDemotions{\roundCounter+1} = \leastDemotions{\roundCounter}+1 = k$, which happens at most once, that we must add another one back to reach the bound in~\eqref{highest-rank-smaller}:
\begin{align}
\demotions{\roundCounter+1}[\demotionCounter] &\leq q\, \demotions{\roundCounter}[\demotionCounter] + 2.\label[ineq]{highest-rank-arrived}
\end{align}

It should be clear that there is nothing the adversary can do to prevent us from obtaining~\invariant{J_\demotionCounter}{\demotionCounter}.
Either he safeguards the leader, in which case $\demotions{\roundCounter}[\demotionCounter]$ quickly reaches zero via~\eqref{highest-rank-fixed}, or he sacrifices the leader until it has $\demotionCounter$~demotions, at which point $\demotions{\roundCounter}[\demotionCounter]$ quickly reaches 1 via~\eqref{highest-rank-past}.

We claim that setting
\begin{align}
 \invariantInc{\demotionCounter} &= \ell_{\demotionCounter} +
      (\leastDemotions{J_\demotionCounter} - \leastDemotions{J_{\demotionCounter-1}}) + 3,\quad\text{with} \\
 \ell_{\demotionCounter} &= \left\lceil\log_\groupSize \demotions{J_{\demotionCounter-1}}[\demotionCounter] \right\rceil
\end{align}
ensures~\invariant{J_\demotionCounter}{\demotionCounter}\ holds for $J_\demotionCounter = J_{\demotionCounter-1}+\invariantInc{k}$.

First, assume that, after $j + 1 = J_{\demotionCounter-1}+i$ rounds, the least-demoted claim stops being demoted before the~$\demotionCounter$th time.
Then,
\begin{align}
\demotions{\roundCounter+1}[\demotionCounter]
    &\leq q^{i} \, \demotions{\roundCounter}[\demotionCounter]+1+\sum_{i'=1}^{i} q^{i'} \\
    &\leq q^{i}\, \demotions{\roundCounter}[\demotionCounter]+2.
\end{align}
From then on, after an additional $i'$ rounds,
\begin{align}
\demotions{\roundCounter+1+i'}[\demotionCounter]
    &\leq q^{i+i'} \, \demotions{\roundCounter}[\demotionCounter]+2 q^{i'}
\end{align}
The value of~\invariantInc{\demotionCounter-1}\ ensures $\demotions{J_\demotionCounter}[\demotionCounter] = 0$.

Conversely, assume that, the least-demoted claim is demoted for the~$\demotionCounter$th time at round $j+1=J_{\demotionCounter-1}+i$,
Then,
\begin{align}
\demotions{\roundCounter+1}[\demotionCounter]
    &\leq q^{i} \, \demotions{\roundCounter}[\demotionCounter]+2+\sum_{i'=1}^{i} q^{i'} \\
    &\leq q^{i}\, \demotions{\roundCounter}[\demotionCounter]+3.
\end{align}
From then on, after an additional $i'$ rounds,
\begin{align}
\demotions{\roundCounter+1+i'}[\demotionCounter]
    &\leq q^{i+i'} \, \demotions{\roundCounter}[\demotionCounter]+3 q^{i'} +p \sum_{i''=1}^{i'} q^{i''} \\
    &\leq q^{i+i'} \demotions{\roundCounter}[\demotionCounter]+3 q^{i'} + 1
\end{align}
The value of~\invariantInc{\demotionCounter-1}\ ensures $\demotions{J_\demotionCounter}[\demotionCounter] \leq 1$.

From~\cref{induction-base}, we know that~$\demotions{J_{\demotionCounter-1}}[\demotionCounter] \le 4\demotionCounter+1$.
We will now find a \emph{constant} bound.
To that end, assume each inductive step takes at least $s \ge 4$ rounds.
Then,
\begin{align}
\demotions{J_{k-1}+s}
        & \lel{k} \boldsymbol{b}*\demotions{J_{k-1}+s-1} + 2\,\boldsymbol{u}_k \\
    & \lel{k} \boldsymbol{b}^2*\demotions{J_{k-1}+s-2} + 2\, \boldsymbol{b}*\boldsymbol{u}_k + 2\, \boldsymbol{u}_k \\
    & \lel{k} \boldsymbol{b}^s*\demotions{J_{k-1}} + 2\, (\boldsymbol{b}^{s-1}+\cdots+\boldsymbol{\delta})*\boldsymbol{u}_k \\
    & \lel{k} \boldsymbol{b}^s*\demotions{J_{k-1}} + 2\, \boldsymbol{s}*\boldsymbol{u}_k,\quad\text{with $\boldsymbol{s}=\boldsymbol{b}^{s-1}+\cdots+\boldsymbol{\delta}$}, \\
\intertext{so that}
\demotions{J_{k-1}+s}
    & \lel{k} \boldsymbol{b}^{2s}*\demotions{J_{k-2}} + 2\,\boldsymbol{s}*\boldsymbol{b}^s*\boldsymbol{u}_{k-1} + 2\,\boldsymbol{s}*\boldsymbol{u}_k \\
    & \lel{k} \boldsymbol{b}^{ks}*\demotions{J_0} + 2\,\boldsymbol{s}*\sum_{j=0}^{k-1} \boldsymbol{b}^{j s} * \boldsymbol{u}_{k-j}  \\
    & \lel{k} \boldsymbol{b}^{ks}*\boldsymbol{r}'_0 + 2s\,\boldsymbol{u}_0 * \sum_{j=0}^{k-1} \boldsymbol{b}^{j s} * \boldsymbol{\delta}_{k-j} \\
    & \lel{k} \boldsymbol{a} + \boldsymbol{a}',
\end{align}
where
\begin{gather}
\boldsymbol{a}=\boldsymbol{b}^{ks}*\boldsymbol{r}'_0\quad\text{and}\quad \boldsymbol{a}'= 2s\,\boldsymbol{u}_0 *  \sum_{j=0}^{k-1} \boldsymbol{b}^{j s} * \boldsymbol{\delta}_{k-j}.
\end{gather}
We are interested in $\demotions{J_{k-1}+s}[k]$.
Taking each term separately,
\begin{align}
\boldsymbol{a}[k]
    &= \sum_{i=0}^{k} \boldsymbol{b}^{ks}[i] \big(4(k-i)+1\big) \\
    &= (4k+1) \sum_{i=0}^{k} \boldsymbol{b}^{ks}[i]\;-\; 4\sum_{i=0}^{k}i\,\boldsymbol{b}^{ks}[i] \\
  \intertext{since $\sum_{i=0}^k i\, \boldsymbol{b}^n[i] = n p \big(\sum_{i=0}^k \boldsymbol{b}^n[i]\big) - (n-k)p\,\boldsymbol{b}^n[k]$,}
    &= 4k(s-1)p\, \boldsymbol{b}^{ks}[k] + \big(4k(1-sp)+1\big) \sum_{i=0}^{k} \boldsymbol{b}^{ks}[i], \label{before-remove-negative} \\
\intertext{so that}
\boldsymbol{a}[k]
    &\le 4k(s-1)p\, \boldsymbol{b}^{ks}[k],\quad\text{since $4k(1-sp)+1<0$.}
\end{align}

We now use the bound~\cref{binomial-bound} again, to write
\begin{align}
\boldsymbol{a}[k]
    &\le 4k(s-1)p\, \tbinom{ks}{k} p^k q^{ks-k} \label{before-removing-binomial}\\
    &\le 4k(s-1)p\, (e s)^k p^k q^{ks-k}. \label{after-removing-binomial}
\end{align}
This attains a maximum at the root of its first derivative
\begin{align}
k=-\big(1+\log (s p q^{s-1})\big)^{-1}
\end{align}
when its second derivative is negative, that is, when
\begin{align}
\frac{G^s}{s} - e (G-1) > 0. \label{second-derivative-negative}
\end{align}
This is not true when $s=3$ and $G=2$, but it is when either $s>3$ and $G\ge2$ or when $s\ge3$ and $G>2$.
To see this, check \eqref{second-derivative-negative} holds for both $s=4, G=2$ and $s=3, G=3$, then verify the derivatives of~\eqref{second-derivative-negative} with respect to~$s$ and~$G$ are both positive when $s\ge3$ and $G\ge2$.
Using $s=4$, we obtain the largest bound for~$\boldsymbol{a}[k] < 5.72$.

To find a bound for~$\boldsymbol{a}'[k]$, note that
\begin{align}
\boldsymbol{a}'[k]
    &=2 s \sum_{\mathclap{i=-\infty}}^{k}\;\, \sum_{j=0}^{k-1} \big(\boldsymbol{b}^{j s} * \boldsymbol{\delta}_{k-j}\big)[i] \\
    &=2 s \sum_{j=0}^{k-1}\;\, \sum_{\mathclap{i=k-j}}^k \boldsymbol{b}^{j s}[i+j-k] \\
    &=2 s \sum_{j=0}^{k-1}\;\, \sum_{i=0}^j \boldsymbol{b}^{j s}[i]. \label{binomial-sum}
\end{align}
To deal with the sum in the form~\mbox{$\sum_{i=0}^{k} b^n[i]$}, we go back to~\eqref{auto-convolution} and recognize the PMF~$B(n,p)$ of a binomial distribution.
Indeed, when~$X\sim B(n,p)$,
\begin{align}
\Pr[X = k] = \boldsymbol{b}^n[k].
\end{align}
We can express the sums in terms of the distribution tail
\begin{align}
    \sum_{i=0}^{k} b^n[i] = \Pr[X \le k],
\end{align}
for which the Azuma-Hoeffding inequality gives the bound
\begin{align}
    \Pr[X \le k] \leq \exp\left( -2n \big(p - \nf{k}{n}\big)^2 \right),\quad\text{for $k\le n p$}.
\end{align}
The condition $k\le n p$ is equivalent to $sp \ge 1$ in~\eqref{binomial-sum}. Substituting,
\begin{align}
\boldsymbol{a}'[k]
    &\le 2 s \sum_{j=0}^{k-1}\;\, \exp\big( -2 j s (p - \nf{1}{s})^2 \big) \\
    &= 2 s \frac{1-\exp\big(-2 k s (p-\nf{1}{s})^2\big)}{1-\exp\big(-2 s (p-\nf{1}{s})^2\big)}  \\
    &< \frac{2 s}{1-\exp \left( -2s (p -\nf{1}{s})^2 \right)} \label{aprimek-bound}.
\end{align}
Since $p \ge \nf{1}{2}$, when $s = 4$ we indeed have $sp \ge 1$, and $p=\nf{1}{2}$ itself maximizes~\eqref{aprimek-bound} to $\boldsymbol{a}'[k] < 20.34$.

This means that
\begin{align}
\demotions{J_{k-1}+s}[k] < 26.05,\quad\text{for all $k > 0$}.
\end{align}

Going back to~\cref{last-round-summation}, we can now write
\begin{align}
J_{\maxDemotions-1} &= J + \sum_{\demotionCounter=0}^{\maxDemotions-1} \invariantInc{\demotionCounter} \\
    &= J + \sum_{\demotionCounter=1}^{\maxDemotions-1} \big(\ell_k + (\leastDemotions{J_\demotionCounter} - \leastDemotions{J_{\demotionCounter-1}}) + 3\big) \\
    &\leq J + (\leastDemotions{J_{\maxDemotions-1}} - \leastDemotions{J_0}) + 3\maxDemotions + \sum_{\demotionCounter=1}^{\maxDemotions-1} \ell_k \\
    &\leq J + 4\maxDemotions + \sum_{\demotionCounter=1}^{\maxDemotions-1} \ell_\demotionCounter,\quad\text{since $0 \leq \leastDemotions{J_{\maxDemotions-1}} \leq \leastDemotions{J_0} \leq \maxDemotions$,} \\
    &\le J + 9\maxDemotions,\quad\text{since $\ell_k < 5$ for all $k$,} \\
    &\le 13 \maxDemotions + \log_\groupSize \nSybils + 2 \sqrt{\maxDemotions \log_\groupSize \nSybils},\quad\text{using $J$ from \eqref{big-J-bound}}.
\end{align}
And this completes the proof.\hfill$\blacksquare$

\bibliographystyle{ACM-Reference-Format}
\bibliography{main}

\end{document}